\documentclass{article}

\PassOptionsToPackage{sort,numbers}{natbib}

\usepackage[preprint]{neurips_data_2024}

\usepackage{longtable}
\usepackage{pdfpages}

\usepackage[utf8]{inputenc} %
\usepackage[T1]{fontenc}    %
\usepackage{hyperref}       %
\usepackage{url}            %
\usepackage{booktabs}       %
\usepackage{amsfonts}       %
\usepackage{nicefrac}       %
\usepackage{microtype}      %
\usepackage{xcolor}         %
\usepackage{graphicx}
\usepackage{array} %
\usepackage{multirow}
\usepackage{subcaption} %
\usepackage{amsmath} %
\usepackage{enumitem}
\usepackage{amssymb}

\newif\ifcomments

\commentstrue

\ifcomments
  
  \newcommand{\chacha}[1]{\textcolor{blue}{[#1 ---\textsc{chacha}]}}

\else
  \newcommand{\chacha}[1]{}
\fi

\usepackage{caption} %

\usepackage{array}
\usepackage{threeparttable}

\usepackage{chngcntr}

\newenvironment{prompt}[2]
  {
    \noindent\begin{minipage}{\textwidth}
    \centering
    \textbf{#1} %
    \vspace{0.5em}
    \begin{tabular}{p{0.1\textwidth} p{0.9\textwidth}}
    \toprule
  }
  {
    \bottomrule
    \end{tabular}
    \end{minipage}
    \vspace{8pt} %
  }

\title{GPT-4V Cannot Generate Radiology Reports Yet}

\author{%
  Yuyang Jiang\thanks{Equal contribution.} \\
  University of Chicago \\
  \texttt{yuyang2001@uchicago.edu} \\
  \And
  Chacha Chen\footnote[1]{} \\
  University of Chicago \\
  \texttt{chacha@uchicago.edu}   %
  \And
  Dang Nguyen \\
   University of Chicago \\
    \texttt{dangnguyen@uchicago.edu } 
    \And
    Benjamin M. Mervak\\
    University of Michigan\\
    \texttt{bmervak@med.umich.edu}
    \And
    Chenhao Tan\\
     University of Chicago \\
      \texttt{chenhao@uchicago.edu}
}

\begin{document}

\maketitle

\begin{abstract}
GPT-4's purported strong multimodal abilities raise interests in using it to automate radiology report writing, but there lacks thorough evaluations.
In this work, we perform a systematic evaluation of GPT-4 (4o and vision-preview) in generating radiology reports across three chest X-ray report benchmarks: MIMIC-CXR, CheXpert Plus, and IU X-Ray.
% on three chest X-ray report benchmarks: MIMIC-CXR, CheXpert Plus, and IU X-Ray. 
We attempt to directly generate reports with different prompting strategies and find that the models fail terribly in both lexical metrics and clinical efficacy metrics. 
To understand the low performance, we decompose the task into two steps: 1) the \textbf{medical image reasoning} step of predicting medical condition labels from images; and 2) the \textbf{report synthesis} step 
of generating reports from (groundtruth) conditions.
% Using GPT-4o as a case study, we show that its performance in image reasoning is consistently low across different prompts.
% In fact, the distributions of model-predicted labels 
% remain constant regardless of groundtruth conditions present on the image,
% suggesting that the model does not interpret chest X-rays meaningfully.
% Even when given groundtruth conditions in report synthesis, its generated reports are less correct and less natural-sounding than a finetuned LLaMA.
% Altogether, our results cast doubt on the viability of using frontier multimodal LLMs in a radiology workflow.
% GPT-4V's purported strong multimodal abilities raise interests in using it to automate radiology report writing, but there lacks thorough evaluations. 
% In this work, we perform a systematic evaluation of GPT-4V in generating radiology reports on two chest X-ray report datasets: MIMIC-CXR and IU X-Ray. 
% We attempt to directly generate reports using GPT-4V through different prompting strategies and find that it fails terribly in both lexical metrics and clinical efficacy metrics. To understand the low performance, we decompose the task into two steps: 1) the medical image reasoning step of predicting medical condition labels from images; and 2) the report synthesis step of generating reports from (groundtruth) conditions.
We show that GPT-4's performance in image reasoning is consistently low across different prompts. In fact, the distributions of model-predicted labels remain constant regardless of which groundtruth conditions are present on the image, suggesting that the model is not interpreting chest X-rays meaningfully. Even when given groundtruth conditions in report synthesis, its generated reports are less correct and less natural-sounding than a finetuned Llama. Altogether, our findings cast doubt on the viability of using GPT-4 in a radiology workflow.

\end{abstract}

\section{Introduction}
\label{sec:intro}

Large language models (LLMs) are becoming multimodal, and GPT-4 model series represent the state-of-the-art.~\footnote{\url{https://huggingface.co/spaces/WildVision/vision-arena}}
Similar to the claimed general-purpose capabilities in LLMs~\citep{bubeck2023sparks, nori2023capabilities}, large multimodal models are supposed to possess advanced skills across a wide range of domains, including high-stakes scenarios such as medicine~\citep{yang2023dawn}.
However, in the field of radiology report generation, where relatively rich datasets are available, there has been \textit{inconclusive} and even \textit{contradictory} evidence regarding the performance of LMMs. 
Some studies~\citep{liu2023holistic,yang2023dawn} claimed that GPT-4V performs well to some extent based on case studies and qualitative analysis. 
In contrast, \citet{Brin2023} found that the model is not yet a reliable tool for radiological image interpretation on a small private dataset.
\citet{wu2023can} observed that GPT-4V can generate structured reports with incorrect content, as evidenced by case studies and qualitative analysis. Moreover, existing evaluation works tend to work with either very small size of samples~\cite{yang2023dawn} or limited evaluation metrics~\cite{li2023comprehensive}.
Our work distinguishes itself by providing an in-depth evaluation and analysis on \textit{why} GPT-4V fails at this task.
\footnote{We access GPT-4 vision model series (inlcuding gpt-4-vision-preview and gpt-4o) through Azure OpenAI service to prevent sharing data with third parties. Due to limited space, we mainly show evaluation results of GPT-4o in the main paper. Throughout the paper, we use GPT-4V to refer to GPT-4o, unless otherwise specified.
Full evaluation results of GPT-4o, GPT-4-vision-preview, and the open sourced Llama3.2-90B-vision-instruct can be found in Appendix~\ref{appen:result}.}

\begin{figure*}[t]
    \centering
    \includegraphics[width=0.85\textwidth]{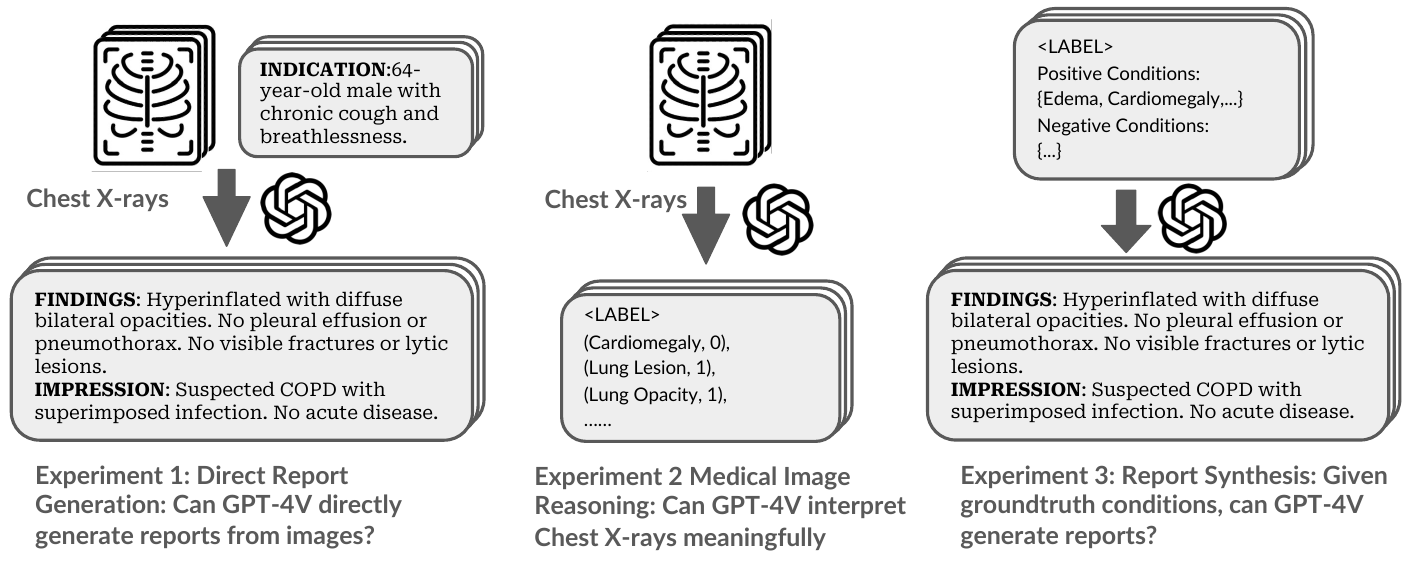}
    \caption{Evaluation overview. In Experiment 1, we evaluate the out-of-box capability of GPT-4V. We further decompose the task into medical image reasoning (Experiment 2) and report synthesis (Experiment 3).
    % \chacha{todo}
    }
    \vspace{-1em}
    \label{fig:exp_overview}
\end{figure*}

To do that, we perform three experiments as shown in Fig.~\ref{fig:exp_overview} on three main radiology report generation benchmarks: MIMIC-CXR, CheXpert Plus, and IU X-Ray.
Our evaluation starts with Experiment 1: \textbf{direct report generation}. 
Different from previous works~\citep{liu2023holistic,yang2023dawn}, we conduct a thorough evaluation of GPT-4V's capability to directly generate reports from chest X-rays, utilizing different prompting strategies and assessing both 
lexical metrics, which measure how textually similar a generated report is to a reference report, and clinical efficacy metrics, which measure how clinically accurate it is.
We experiment with various prompting strategies, including zero-shot, contextual enhancement, chain-of-thought (CoT) \citep{wei2022chain}, and few-shot in-context learning. 
Despite our various attempts,
the performance of GPT-4V is consistently low in both 
metrics.

To further investigate the reason for GPT-4V's poor performance, we break down  
report generation into two steps, \textbf{medical image reasoning} and \textbf{report synthesis given medical conditions}.
For Experiment 2 (medical image reasoning), we first test whether GPT-4V can identify medical conditions from X-rays. 
Our findings indicate that GPT-4’s performance in identifying medical conditions from images is unsatisfactory across different prompts. 
Based on limited capability results, we further 
compare the difference between distributions of predicted medical condition labels conditioned on different groundtruth image labels.
We find that GPT-4V cannot interpret medical images meaningfully as the distribution of predicted labels does not vary depend on the groundtruth label. 

Finally, in Experiment 3 (report synthesis), we explore whether bypassing the image reasoning bottleneck by providing groundtruth conditions enables GPT-4V to generate clinically usable reports. 
As expected, reports generated by GPT-4V achieve higher clinical efficacy; 
however, the limited improvement in lexical metrics suggests that GPT-4V-generated reports 
remain dissimilar to human-written reports in style.
Most importantly, GPT-4V underperforms a finetuned Llama-2 in both lexical metrics and clinical efficacy metrics, calling into question its utility.
We further validate our findings by conducting an additional human reader study with two radiologist to evaluate the clinical viability of GPT-4V generated reports. 

In summary, our key contributions and conclusions are as follows:

\begin{itemize}[leftmargin=*,topsep=-2pt,itemsep=-2pt]
    \item We perform the first systematic and in-depth evaluation to benchmark GPT-4V in radiology report generation.
    Our main conclusion is that GPT-4V cannot generate radiology reports yet. 
    \item To understand the poor performance, we decompose the task into medical image reasoning and report synthesis. We find that GPT-4V cannot interpret chest X-ray images meaningfully in the image reasoning step, and further validate this finding through rigorous hypothesis testing.
    \item During report synthesis, we address the image reasoning bottleneck by providing groundtruth conditions. Nonetheless, both experimental results and human evaluations consistently show that GPT-4V performs worse than a finetuned Llama-2 baseline.
\end{itemize}

\section{Related Work}
\label{sec:literature}
While there is an emerging line of work in investigating the direct application of GPT-4\footnote{GPT-4V mentioned in previous literature typically refers to GPT-4-vision-preview.} in radiology report generation, 
there lacks a systematic evaluation. 
\cite{liu2023holistic, wu2023can, yang2023dawn} tested capabilities for general medical applications through case studies, including selected examples of chest X-ray reports with qualitative analysis.
\cite{Brin2023} provided quantitative results on GPT-4V's accuracy in interpreting medical images, using a small private dataset that includes chest X-rays.
But their evaluation only focused on identifying the imaging modality (e.g. CT, ultrasound, or MRI) and the anatomical region of the pathology, rather than assessing the overall quality of generated radiology reports.
\cite{li2023comprehensive} evaluated GPT-4V on the public \textsc{mimic-cxr} dataset, but only used lexical and semantic metrics without assessing clinical efficacy.
\cite{chaves2024clinically} included GPT-4V as one of the baselines. However, their focus is on proposing a new model. 
In contrast, we provide an in-depth evaluation across various metrics with different prompting strategies on two public datasets.

Prior work has also examined text-only applications of GPT-4 related to radiology report generation, such as summarizing findings~\citep{liu2024large,sun2023evaluating}, handling various text processing tasks including sentence semantics, structural extraction, and summary of findings~\citep{liu2023exploring}, radiology board-style exams~\citep{bhayana2023gpt}, detecting errors in radiology reports~\citep{gertz2024potential}, and refining human-written reports for better standardization and clarity~\citep{hasani2023evaluating}.
Additionally, other related multimodal tasks include visual question answering based on radiology images~\citep{yan2023multimodal} and biomedical image classification~\citep{liu2023holistic}.

To the best of our knowledge, our work provides the first systematic and in-depth evaluation of GPT-4V's capabilities to generate radiology reports.

\section{Experiment Setup}
\label{sec:exp}

\begin{table*}
  \centering
    \resizebox{.95\textwidth}{!}{
  \begin{tabular}{ll}
    \toprule
    \multicolumn{2}{l}{\textbf{Experiment 1: Direct Report Generation}}\\
    \midrule
    \hyperref[prompt1.1]{Prompt 1.1} Basic generation   & Direct report generation based on chest X-ray images \\
    \hyperref[prompt1.2]{Prompt 1.2} +Indication  &  Contextual enhancement by providing the indication section\\
    \hyperref[prompt1.3]{Prompt 1.3} +Instruction & Contextual enhancement by providing instructions on medical conditions\\
   \hyperref[prompt1.4]{Prompt 1.4} Chain-of-Thought (CoT)  & 
    Step 1 - medical condition labeling; Step 2 - report synthesis \\
    \hyperref[prompt1.5-mimic]{Prompt 1.5} Few-shot  & Few-shot: in-context learning given a few examples\\
    \midrule
    \multicolumn{2}{l}{\textbf{Experiment 2: Medical Image Reasoning Capability}}\\
    \midrule
   \hyperref[prompt2.1]{Prompt 2.1} Image reasoning  & Medical condition labeling directly from chest X-ray images  \\
    \midrule
    \multicolumn{2}{l}{\textbf{Experiment 3: Report Synthesis Given Medical Conditions}}\\
    \midrule
    \hyperref[prompt3.1]{Prompt 3.1} Report synthesis  & Report generation using provided positive and negative conditions \\ 
    \bottomrule
  \end{tabular}}
\caption{An index to prompts used in all of our experiments.}
  \label{tab:sum}
\end{table*}

\paragraph{Method.}
In Experiment 1 (Section~\ref{sec:exp1}), we evaluate GPT-4V's 
ability to directly generate
radiology report given chest X-ray images. 
We consider five variations of prompts as outlined in Table~\ref{tab:sum}.
Prompt 1.1 (Basic generation)
is a prompt to test the out-of-the-box capability of GPT-4V. 
We implement three additional prompting strategies leveraging insights in prompt engineering: (1) 
inspired by \cite{nguyen2023pragmatic}, we add relevant contextual information (i.e., the \textsc{indication}) to derive Prompt 1.2 as ``Indication enhancement'', and add 
instructions on medical condition labels to Prompt 1.3 as ``+instruction'' enhancement; (2) we use a chain-of-thought (CoT) strategy in Prompt 1.4, eliciting the model with two steps: medical condition label prediction based on images followed by report synthesis based on the predicted labels;
(3) We adopt few-shot in-context learning by adding a few example image-report pairs in Prompt 1.5. 
We compare these results with the state-of-the-art (SOTA) models. 

In addition to evaluation of the end-to-end radiology report generation capability, we further evaluate on the decomposed tasks: Experiment 2 (Section~\ref{sec:exp2}): chest X-ray image reasoning; and Experiment 3 (Section~\ref{sec:exp3}): synthesizing a radiology report from given conditions.
This decomposition allows us to look into the bottlenecks in the current generation performance.
In Experiment 2, we prompt the model to directly output medical condition labels from images (Prompt 2.1). 
In Experiment 3, we bypass image reasoning to test GPT-4V's textual synthesis ability and provide groundtruth conditions to evaluate the model's report composition capability independently (Prompt 3.1).
To contextualize the performance of GPT-4V, we also report the performance of a finetuned Llama-2-7B on groundtruth labels and groundtruth impressions following Alpaca~\citep{taori2023alpaca}.

\paragraph{Dataset and pre-processing.}
We use three chest X-ray datasets: {MIMIC-CXR}, {IU X-ray} and {CheXpert Plus}. 
The {MIMIC-CXR} dataset~\citep{johnson2019mimic} contains chest X-ray images and their corresponding free-text radiology reports. The dataset includes 377,110 images from 227,835 studies. Each study has one radiology report and one or more chest X-rays. 
The {IU X-ray}dataset~\citep{demner2016preparing}  (also known as ``Open-i'')  includes 3996 de-identified radiology reports and 8121 associated images from the Indiana University hospital network.
{CheXpert Plus} dataset~\citep{chambon2024chexpertplusaugmentinglarge} is a newly-released and comprehensive radiology data collection, featuring a total of 223,462 unique pairs of radiology reports and chest X-rays across 187,711 studies from 64,725 patients.
For our evaluation, we use the whole validation set of {CheXpert Plus} (200 samples) and randomly sample 300 studies from each of the {MIMIC-CXR} and {IU X-RAY} datasets after removing studies with empty impression or indication sections.
This results in a total of 800 samples.
More details about data processing can be checked in Appendix \ref{appen:data}.

\begin{table*}[t]
\centering
\resizebox{\textwidth}{!}{
    \begin{threeparttable}
\begin{tabular}{@{}p{2cm}@{\hspace{8pt}}p{1.5cm}@{\hspace{2pt}}p{1.9cm}@{\hspace{2pt}}p{1.5cm}@{\hspace{2pt}}p{1.9cm}@{\hspace{2pt}}p{1.8cm}@{\hspace{2pt}}p{1.8cm}@{\hspace{2pt}}p{1.8cm}@{\hspace{2pt}}p{2cm}@{\hspace{2pt}}p{2cm}@{\hspace{2pt}}p{2cm}@{}}
\toprule
\multirow{2}{*}{\textbf{Experiment}} & \multicolumn{4}{c}{\textbf{Lexical metrics}} & \multicolumn{6}{c}{\textbf{Clinical Efficacy Metrics}} \\
\cmidrule(lr){2-5} \cmidrule(l){6-11}
& \textbf{BLEU-1} & \textbf{BLEU-4} & \textbf{ROUGE} & \textbf{METEOR} & \textbf{Pos F1} & \textbf{Pos F1@5} & \textbf{Rad. F1} & \textbf{Neg F1$^*$} & \textbf{Neg F1@5$^*$} & \textbf{Hall.$^*$}$\downarrow$ \\
\midrule
\multicolumn{11}{c}{\textsc{MIMIC-CXR}} \\
Basic & 0.331 & 0.031 & 0.225 & 0.282 & 0.134 & 0.153 & 0.163 & 0.038 & 0.081 & 0.587 \\
+Indication & 0.333 & 0.044 & 0.234 & 0.290 & \bf 0.273 & 0.255 & 0.182 & \bf 0.048 & \bf 0.126 & 0.547 \\
+Instruction & 0.276 & 0.028 & 0.203 & 0.264 & 0.155 & 0.230 & 0.154 & 0.042 & 0.101 & \bf 0.359 \\
CoT & 0.242 & 0.017 & 0.186 & 0.201 & 0.172 & \bf 0.272 & 0.119 & 0.008 & 0.020 & 0.397 \\
Few-shot & \bf 0.337 & \bf 0.055 & \bf 0.257 & \bf 0.301 & 0.170 & 0.203 & \bf 0.188 & 0.035 & 0.091 & 0.477 \\
\hline
Llama-3.2$\textsuperscript{1}$ & 0.258 & 0.026 & 0.249 & 0.216 & 0.172 & 0.300 &0.144 & 0.045 & 0.117 & 0.049\\
\hline
SOTA \newline [ref.] & 0.402 \newline ~\citeauthor{liu2024bootstrapping} & 0.142 \newline~\citeauthor{hyland2023maira} & 0.291 \newline~\citeauthor{liu2024bootstrapping} & 0.333 \newline~\citeauthor{hyland2023maira} & 0.473 \newline~\citeauthor{liu2024bootstrapping} & 0.516 \newline~\citeauthor{tu2024towards} & 0.267 \newline~\citeauthor{tu2024towards} & 0.077 \newline~\citeauthor{nguyen2023pragmatic} & 0.156 \newline~\citeauthor{nguyen2023pragmatic} & 0.158 \newline~\citeauthor{nguyen2023pragmatic} \\
\midrule
\(\Delta\)(to SOTA) & -16.17\% & -61.27\% & -11.68\% & -9.61\% & -42.28\% & -47.29\% & -29.59\% & -37.66\% & -19.23\% & 20.10\% \\
\midrule
\multicolumn{11}{c}{\textsc{IU X-Ray}} \\
Basic & 0.316 & 0.045 & 0.238 & 0.311 & 0.059 & 0.045 & 0.203 & 0.000 & 0.000 & 0.303 \\
+Indication & \bf 0.330 & \bf 0.049 & 0.242 & \bf 0.323 & 0.077 & 0.098 & \bf 0.214 & \bf 0.071 & \bf 0.051 & 0.307 \\
+Instruction & 0.238 & 0.030 & 0.207 & 0.283 & \bf 0.081 & 0.146 & 0.174 & 0.000 & 0.000 & \bf 0.177 \\
CoT & 0.239 & 0.024 & 0.194 & 0.231 & 0.077 & \bf 0.161 & 0.144 & 0.000 & 0.000 & 0.197 \\
Few-Shot & 0.279 & 0.044 & \bf 0.243 & 0.250 & 0.037 & 0.031 & 0.187 & 0.010 & 0.025 & 0.211 \\
\hline
Llama-3.2 & 0.248 & 0.027 & 0.239 & 0.231 & 0.104 & 0.214 & 0.158 & 0.015 & 0.040 & 0.011 \\
\hline
SOTA\newline [ref.] & 0.499 \newline~\citeauthor{liu2024bootstrapping} & 0.184 \newline~\citeauthor{liu2024bootstrapping} & 0.390 \newline~\citeauthor{liu2024bootstrapping} & 0.208 \newline~\citeauthor{liu2024bootstrapping} & - & - & - & - & - & - \\
\midrule
\(\Delta\)(to SOTA) & -36.27\% & -75.82\% & -38.72\% & +49.52\% & - & - & - & - & - & - \\
\midrule
\multicolumn{11}{c}{\textsc{CheXpert Plus}} \\
Basic & \bf 0.237 & \bf 0.015 & \bf 0.176 & \bf 0.191 & 0.228 & 0.191 & \bf 0.112 & 0.013 & 0.035 & 0.680 \\
+Instruction & 0.191 & 0.007 & 0.159 & 0.172 & 0.210 & 0.325 & 0.101 & 0.042 & 0.085 & 0.377 \\
CoT & 0.166 & 0.011 & 0.155 & 0.139 & \bf 0.234 & \bf 0.339 & 0.077 & 0.000 & 0.000 & 0.400 \\
Few-shot & 0.171 & 0.007 & 0.158 & 0.149 & 0.188 & 0.224 & 0.094 & \bf 0.043 & \bf 0.111 & \bf 0.370 \\
\hline
Llama-3.2 & 0.166 & 0.006 & 0.175 & 0.147 & 0.261 & 0.355 & 0.092 & 0.031 & 0.081 & 0.058 \\
\hline
SOTA  & - & 0.069  & 0.279  & - & 0.366  & 0.495  & 0.285  & - & - & - \\
\text{[ref.]}& & \multicolumn{3}{l}{\citeauthor{chambon2024chexpertplusaugmentinglarge} } &
\multicolumn{3}{c}{\citeauthor{chambon2024chexpertplusaugmentinglarge} }\\
\midrule
$\Delta$(to SOTA) & - & -78.26\% & -36.92\% & - & -36.07\% & -31.52\% & -60.70\% & - & - & - \\
\bottomrule
\end{tabular}
\begin{tablenotes}
\item $^*$ To compare with SOTA numbers, all metrics, except for those marked with $^*$ (Neg F1, Neg F1@5, and Hall), are evaluated on the findings section. $^*$ columns are based on the impression section.
\item CheXpert Plus doesn't have indication section in reports, thus we skip experiment with +Indication prompt.
\item $\textsuperscript{1}$ Due to the space limit, we only show the best results of prompt 1.1-1.5 for Llama-3.2-90B-Vision-Instruct. It is noted that the low hallucination rate is likely because it only outputs medical conditions (mostly wrong but not uninferable). 
\item Full details of the performance of GPT-4o, GPT-4-vision-preview, and Llama3.2-90B-Vision-Instruct, including results for both the findings and impression sections, are provided in the Appendix~\ref{appen:result}.
\end{tablenotes}
\end{threeparttable}
}
\caption{Direct report generation performance comparison. GPT-4V shows a significant performance gap compared to SOTA, and the results are consistent across the five prompting strategies. Open sourced Llama3.2 performs similarly compared with GPT-4V. Examples of generated reports across different prompts can be found in Appendix~\ref{appen:reports}.}
\label{tab:gpt4o-exp1}
\end{table*}

\paragraph{Evaluation metrics.}
We evaluate the generated reports from two aspects: 

\begin{itemize}[leftmargin=*,topsep=0pt]
\item Lexical metrics. Lexical metrics focus on the surface form and the exact word matches between the generated and reference texts. We adopt common lexical metrics: BLEU~\citep{papineni2002bleu} ($1$-gram and $4$-gram), ROUGE-L~\citep{lin2004rouge}, and METEOR~\citep{banerjee2005meteor}.

\item Clinical efficacy metrics. 
We first evaluate on \textbf{clinical correctness} based on labeler results on generated reports. 
Following existing works~\citep{hyland2023maira,tu2024towards,nguyen2023pragmatic}, we use the CheXbert automatic labeler~\citep{smit2020chexbert} to extract labels for each of 14 Chexpert medical conditions~\citep{irvin2019chexpert}. We compute both positive F1 and negative F1, where each condition has four labels: \texttt{{present, absent, uncertain, unmentioned}}. Positive F1 considers only positive labels against all others, while negative F1 considers negative labels as 1 and all other labels as 0.
We report the macro-averaged F1 on all 14 conditions and on top 5 conditions (which  only reports on the five most common conditions\footnote{Top five conditions are Pneumothorax, Pneumonia, Edema, Pleural Effusion, and Consolidation.}). We also report RadGraph F1~\citep{jain2021radgraph}, which captures the overlap in clinical entities and relations between a generated report and a reference report.

Additionally, from a \textbf{pragmatic} viewpoint, commenting on negative observations is essential in radiology reports.
Following \citet{nguyen2023pragmatic}, we compute Negative F1 and Negative F1-5, to evaluate whether the model can 
accurately identify negative conditions and include that in the generated reports. All F1 scores are macro-averaged. 
We also use the \textbf{hallucination} metric to quantify the proportion of uninferable information. Following \citet{nguyen2023pragmatic}, we define uninferable information to include previous studies, previous treatment details, recommendations, doctor communications, and image view descriptions.
\end{itemize}

\section{Results}
\label{sec:result}

\subsection{Experiment 1: Can GPT-4V directly generate reports from images?}
\label{sec:exp1}

We first evaluate the out-of-the-box capability of GPT-4V in generating radiology reports from chest X-ray images using basic generation (\hyperref[prompt1.1]{Prompt 1.1}). 
Table \ref{tab:gpt4o-exp1} shows the results compared with existing state-of-the-art (SOTA) models. 
Overall, GPT-4V significantly underperforms the state-of-the-art models on both lexical and clinical efficacy metrics, with the exception of the METEOR score on the \textsc{IU X-ray} dataset. 
The relatively better METEOR performance is due to its comprehensive evaluation criteria, which include synonymy and paraphrasing, not just exact word matches like BLEU and ROUGE. This allows METEOR to recognize semantic equivalents, even if the word choice differs. In other words, the generated report somewhat resembles a radiology report, although it fails at the exact word-level matching.
For clinical efficacy metrics, the gaps to SOTA are consistently large.
This suggests that GPT-4V struggles to accurately identify conditions in its generated reports from images alone.

\paragraph{Our results are consistent across prompting strategies.}
Our prompting strategies include adding contextual information, chain-of-thought reasoning, and few-shot prompting. 
While indication enhancement (\hyperref[prompt1.2]{Prompt 1.2}) provides indication section as input in addition to chest X-rays and 
improves many metrics for both \textsc{MIMIC-CXR} and \textsc{IU X-ray}, it remains within the same range and does not significantly reduce the gap compared to SOTA. 
Instruction enhancement (\hyperref[prompt1.3]{Prompt 1.3}) provides medical condition descriptions and makes a moderate yet still limited difference to SOTA in Positive F1 scores and Hallucination.
Following the same labeling instructions, Chain-of-Thought (\hyperref[prompt1.4]{Prompt 1.4}) similarly increases Positive F1-5 by $11.9\%$ in {MIMIC-CXR}, $11.6\%$ in \textsc{IU X-ray} and $14.8\%$ in {CheXpert Plus}, marking the most effective advances so far. However, it still faces a substantial gap to SOTA, with $47.29\%$ in {MIMIC-CXR} and $31.52\%$ in {CheXpert Plus}.
Few-Shot (\hyperref[prompt1.5-mimic]{Prompt 1.5}) provides image-report pairs as context and generally improves lexical metrics, RadGraph F1, and Hallucination, while clinical correctness, particularly in identifying positive conditions, remains consistently low across three datasets. This indicates that while few-shot prompting might help GPT-4V mimic the format of groundtruth reports, it still falls short in generating accurate reports.

\subsection{Experiment 2: Can GPT-4V interpret chest X-rays meaningfully?}
\label{sec:exp2}

\begin{table}[t]
\centering
\footnotesize
% \resizebox{\columnwidth}{!}{
\begin{tabular}{lccccc}
\toprule
\textbf{Metric} & \textbf{Chain-of-Thought (1st Step)} & \textbf{Image Reasoning} \\
\midrule
\textbf{Positive F1} & 0.195 & 0.161 \\
\textbf{Positive F1@5} & 0.298 & 0.242 \\
\bottomrule
\end{tabular}
% }
\caption{
Image reasoning performance of GPT-4V on MIMIC-CXR. 
The model performs poorly in identifying medical conditions from chest X-ray images.
% The results show positive F1 scores for correctly predicting the presence of medical conditions.
}
\label{tab:gpt-4o-exp2}
\end{table}

\begin{table}[t]
\centering
  \small
  % \resizebox{\columnwidth}{!}{
  \begin{tabular}{ccccc}
    \toprule
    \multirow{2}{*}{\bf Statistics} & \multicolumn{2}{c}{\bf Overall} & \multicolumn{2}{c}{\bf Top 5 Conditions} \\
    \cmidrule(r){2-3}
    \cmidrule(l){4-5}
    & \bf Groundtruth & \bf GPT-4o & \bf Groundtruth & \bf GPT-4o \\
    \midrule
    $\chi^2$ statistic & 1770.38 & 66.05 & 243.51 & 5.30 \\
    p-value & p < 1e-4 & 1.00 & p < 1e-4 & 0.994 \\
    df. & 144 & 144 & 16 & 16 \\
  \bottomrule
\end{tabular}
% }
  \caption{$\chi^2$-test for homogeneity of label distribution across different condition groups. When p-value is smaller than 0.0001,  at 0.01\% significance level, we can reject the null hypothesis that different groups follow the same label distribution.}
    \label{tab:gpt-4o-exp2-chi}
\end{table}

\begin{figure*}[htbp]
  \centering
  \includegraphics[width=\textwidth]{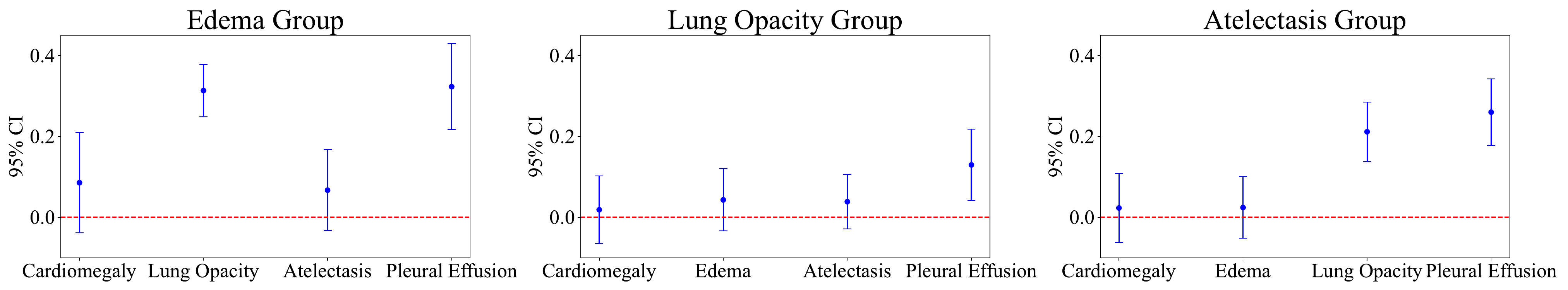}
  \captionsetup{skip=2pt}
  \caption{95\% Bootstrap confidence interval of three example conditions for MIMIC-CXR. When zero falls into the interval, at 95\% confidence level, we cannot reject the null hypothesis that GPT-4V labels $j$-th condition independent of which condition group this study belongs to.}
  \label{fig:ci}
\end{figure*}

In this section, we probe GPT-4V's ability to reason about chest X-ray images alone.
Specifically, we evaluate whether the model can meaningfully interpret chest X-ray images by measuring how accurately GPT-4V can label medical conditions present (positive F1). 
Table ~\ref{tab:gpt-4o-exp2} provides an overview of GPT-4V's labeling performance under different prompting strategies.

We can see that GPT-4V cannot accurately specify positive conditions from given chest X-rays. This can be highlighted by consistently poor Positive F1 scores observed from various prompting strategies. Furthermore, this inability to accurately interpret images may directly contribute to GPT-4V's failure in generating high-quality reports, as supported by similar Positive F1 score of 0.172 and Positive F1-5 score of 0.272 from the report synthesis phase of Chain-of-Thought (see Table~\ref{tab:gpt4o-exp1}), compared to 0.195 (Positive F1) and 0.298 (Positive F1-5) from the initial label generation phase of Chain-of-Thought.

Overall, these results indicate GPT-4V's limited ability in identifying medical conditions from chest X-ray images, regardless of whether labels are derived from CoT 1st step or 
direct prompting.

\paragraph{Testing 
whether GPT-4V generates labels based on given chest X-rays.}
\label{sec:exp3}

Considering the failure of GPT-4V to accurately label medical conditions, we would like to investigate to what extent can GPT-4V predict meaningful labels given a specific chest X-ray image.
To test this, we group chest X-rays by their groundtruth conditions and then analyze the generated label distribution for each group. 
If the label distributions are similar across different condition groups, it would suggest that GPT-4V is not meaningfully identifying labels from the chest X-rays but rather assigning labels randomly without proper image interpretation. 
For example, if the model's generated label probabilities are roughly the same regardless of whether the groundtruth condition of the given image is Edema or Cardiomegaly, it indicates a limited capability in medical image reasoning.

Formally, let $X_{ij}$ be a binary random variable that takes the value 1 if GPT-4V labels the $j$-th condition as positive for the chest X-ray image associated with the $i$-th study, and 0 otherwise, where $i = 1, 2, \ldots, 300$ and $j = 1, 2, \ldots, 13$. We exclude the ``No Findings'' condition from this study.
We define $Y_{j}=\sum_{i=1}^{300}X_{ij}$ as the sum of positive mentions for the $j$-th condition across all 300 studies, and $\mathbf{Y} = [Y_1, \ldots, Y_{13}]$ as the count vector. 
Next, we categorize the study pool into 13 condition groups, where group $k$ consists all studies that are ground truth positive for the $k$-th condition based on the associated radiology report. Note that there might be overlaps between these groups, as a single study can be positive for multiple conditions.
For each group $k$, GPT-4V's labeling process given the chest X-ray image from $i$-th study can be modeled as:

\begin{equation}
\normalsize
\left\{ 
\begin{array}{l} 
X_{ij}^{(k)} \sim \text{Bernoulli}(P_j^{(k)}) \quad \text{ for } i \in \text{group } k \text{ and } j = 1, \ldots, 13\\
\mathbf{Y}_k \sim \text{Multinomial}(n_k; \mathbf{P}_k) \quad \text{ with } \mathbf{P}_k = [P_1^{(k)}, \ldots, P_{13}^{(k)}] 
\end{array} 
\right.
\end{equation}

where $n_k$ is the number of studies in group $k$, and $P_j^{(k)}$ is the probability that GPT-4V labels the $j$-th condition as positive for the chest X-ray images associated with the studies in group $k$.

\begin{table*}[htbp]
\centering
\resizebox{\textwidth}{!}{
\begin{threeparttable}
\begin{tabular}{lcccccccccc}
\toprule
\multirow{2}{*}{\textbf{Experiment}} & \multicolumn{4}{c}{\textbf{Lexical metrics}} & \multicolumn{6}{c}{\textbf{Clinic Efficacy Metrics}} \\
\cmidrule(lr){2-5} \cmidrule(l){6-11}
& \textbf{BLEU-1} & \textbf{BLEU-4} & \textbf{ROUGE} & \textbf{METEOR} & \textbf{Pos F1} & \textbf{Pos F1@5} & \textbf{Rad. F1} & \textbf{Neg F1} & \textbf{Neg F1@5} & \textbf{Hall.}$\downarrow$ \\
\midrule
GPT-4o & 0.159 & 0.006 & 0.142 & 0.185 & 0.123 & 0.152 & 0.077 & 0.038 & 0.081 & 0.587 \\
GPT-4o (gt) & 0.175 & 0.009 & 0.187 & 0.183 & 0.879 & 0.972 & 0.105 & 0.639 & 0.956 & \textbf{0.427} \\
GPT-4-vision-preview (gt) & 0.176 & 0.007 & 0.185 & 0.179 & 0.885 & \bf 0.977 & 0.103 & 0.584 & \bf 0.958& 0.431\\
Finetuned Llama-2 (gt) & \bf 0.301 & \bf 0.094 & \bf 0.330& \bf 0.348& \bf 0.923 & 0.957& \bf 0.286& \bf 0.703&  0.941&  0.710\\
\bottomrule
\end{tabular}
\end{threeparttable}
}
\caption{Performance in report generation with groundtruth conditions. Although GPT-4V's performance improves significantly, it still underperforms finetuned Llama-2, especially in matching the writing style of groundtruth reports.
}
\label{tab:gpt-4o-exp3}
\end{table*}

\begin{table*}[htbp]
\centering
\small
    \resizebox{\textwidth}{!}{
\begin{threeparttable}
\begin{tabular}{lccccc}
\toprule
 & \textbf{Binary} & \multicolumn{3}{c}{\textbf{Likert Scale (1-5)}} \\
\cmidrule(lr){2-2} \cmidrule(lr){3-5}
& \textbf{Clinically Usable} & \textbf{Diagnostic Accuracy} & \textbf{Completeness} & \textbf{Clarity/Readability} \\
\midrule
Groundtruth & 50/50 (100\%) & \bf 4.86 & \bf 4.87 & 4.92 \\
Finetuned Llama-2 & 46/50 (92\%) & 4.16 & 4.42 & \bf 4.94 \\
GPT-4-vision-preview & 36/50 (72\%) & 3.63 & 3.49 & 3.80 \\
\bottomrule
\end{tabular}
\begin{tablenotes}
\item All metrics are evaluated on the impression section and are averaged based on results from two experts.
\end{tablenotes}
\end{threeparttable}
}
\caption{Human evaluation of radiology report quality. From the perspective of radiologists, GPT-4V underperforms relative to the finetuned Llama-2, particularly in fine-grained likert scale metrics of diagnostic accuracy, completeness, and clartiy/readability.}
\label{tab:gpt-4v-exp3-hm}
\end{table*}

We first use  a \textbf{$\chi^2$-test} to test if GPT-4V follows the same label distribution across different groups, i.e., testing the null hypothesis (\(H_0\)) that \(\mathbf{P}_{k} = \mathbf{P}_{k'} \) for any groups \(k\) and \(k'\). 
Additionally, we use \textbf{bootstrap confidence interval}~\citep{davison1997bootstrap} to test if GPT-4V labels one certain condition independently of the groundtruth condition group. 
Specifically, we test the null hypothesis (\(H_0\)) that \(P_{j}^{(k)} = P_{j}\) for any condition \(j\) and group \(k\). 
More test details and robustness check can be found in Appendix~\ref{appen:test}.

Table~\ref{tab:gpt-4o-exp2-chi} presents $\chi^2$-test results for the homogeneity of label distribution across different groups. For both the overall and top 5 conditions\footnote{Due to the sparsity of the original study pool, we report results for two different tables: (1) A modified table with zero elements replaced by 0.001; (2) A reduced table with only five most frequent medical conditions in the subsample.}, at 0.01\% significance level, we can both reject the null hypothesis for groundtruth reports that different groups follow the same label distribution, but not for GPT-4V's generated reports.

Figure~\ref{fig:ci} illustrates the 95\% bootstrap confidence intervals for three example conditions. If zero falls within the interval, we cannot reject the null hypothesis that GPT-4V labels the $j$-th condition independently of the condition group at 95\% confidence level. 
The figure shows that, in 7 out of 12 cases (58.3\%), we cannot reject the null hypothesis.

In summary, the results show that GPT-4V labels conditions independently of the groundtruth condition, and there is no significant difference in label distributions across groups in GPT-4V's generated reports, unlike the groundtruth reports.

\subsection{Experiment 3: Given groundtruth conditions, can GPT-4V generate reports?}
\label{sec:exp4}

Given that GPT-4V cannot perform image reasoning, we investigate whether GPT-4V can 
generate high-quality radiology reports given groundtruth medical conditions. We conduct an experiment on report synthesis (\hyperref[prompt3.1]{Prompt 3.1}) and use a finetuned Llama-2 model as a baseline for comparison.

Table~\ref{tab:gpt-4o-exp3} shows that while using groundtruth conditions significantly enhances GPT-4V's clinical accuracy, it still does not perform as well as the finetuned Llama-2, particularly in matching the content of groundtruth reports. 
Progress in clinical accuracy is evidenced by large improvements in F1 scores compared to basic generation (\hyperref[prompt1.1]{Prompt 1.1}). However, there are only minor changes in lexical metrics and RadGraph F1, which focus on entity relation matching in groundtruth reports, along with consistently large gaps with finetuned Llama-2, suggest that groundtruth conditions are insufficient to align GPT-4V's writing closely with that of groundtruth reports. 
The higher scores of the finetuned Llama-2 in lexical metrics also indicate that finetuning open models is an effective way to leverage existing datasets.

\begin{table*}[t]
\centering
\small %
\begin{tabular}{@{}p{2.05cm}p{1.35cm}p{5.5cm}p{0.7cm}p{0.3cm}p{0.7cm}p{0.7cm}@{}}
\toprule
\bf Groundtruth labels & \textbf{Model} & \textbf{Report} & \textbf{Usable} & \textbf{Acc.} & \textbf{Comp.} & \textbf{Clarity} \\
\midrule
\multirow{3}{2.1cm}{%
Positive: Enlarged Cardiomediastinum, Lung Opacity, Atelectasis \newline \newline Negative: Pneumonia} & Groundtruth & In comparison with study of \_\_\_, there is little change in the cardiomediastinal silhouette and pacer leads. Continued elevation of the right hemidiaphragmatic contour. Opacification medially above the elevated hemidiaphragm most likely represents atelectatic changes. No definite acute focal pneumonia. \newline  & \checkmark & 5 & 5 & 4 \\
 & Finetuned Llama-2 & 1. No evidence of pneumonia. 2. Stable postoperative appearance of the chest. 3. Increased retrocardiac opacity likely reflects atelectasis. \newline & \checkmark & 5 & 5 & 5 \\
 & GPT-4-vision-preview & The chest radiograph demonstrates an enlarged cardiomediastinum, lung opacity which may be due to various etiologies, and atelectasis. There is no clear evidence of pneumonia on the radiograph.  & \checkmark & 3 & 4 & 3 \\
\bottomrule
\end{tabular}
\caption{Comparison of example reports from human annotation.}
\label{tab:gpt-4v-hm-example}
\end{table*}

\paragraph{Human Evaluation}
\label{sec:human}

To further evaluate the quality of GPT-4V-generated reports beyond automatic metrics, we collaborate with two experts: one board-certified and one board-eligible radiologist, to conduct a human evaluation.
From our testing set of 300 studies, we randomly select 50 cases for blind human evaluation. 
Each radiologist is provided with anonymized chest X-ray images and randomly ordered \textsc{impression} sections from groundtruth reports, as well as reports generated by Llama-2 and GPT-4V.
Both Llama-2 and GPT-4V are prompted with groundtruth medical conditions. 
The evaluation involves a detailed review of three reports per study case, assessing each report's clinical usability with a binary label as the first step.
Then, the radiologist rates each report on two dimensions: clinical efficacy (diagnostic accuracy and completeness) and lexical performance (clarity/readability). 
Reports are rated on a Likert scale, where a score of 5 denotes superior performance and a score of 1 denotes poor performance. 
We compute and report the average scores for each metric across different report types.

Table~\ref{tab:gpt-4v-exp3-hm} shows that, from the perspective of radiologists, GPT-4V still underperforms the finetuned Llama-2.
Groundtruth reports are indeed of high quality, rated as clinically usable in 50 out of 50 cases.
However, a significant usability gap is observed between Llama-2 and GPT-4V, with Llama-2 being deemed clinically usable in 46 out of 50 cases, compared to 36 out of 50 for GPT-4V.
Furthermore, Llama-2 consistently outperforms GPT-4V across all other Likert scale metrics, especially in completeness and clarity/readability. 

Table~\ref{tab:gpt-4v-hm-example} presents three example reports. While groundtruth reports offer detailed clinical insights and varied descriptors, GPT-4V tends to provide vague statements, only stating ``lung opacity which may be due to various etiologies'' without specifying its location, severity, or offering a differential diagnosis. Llama-2 performs slightly better by offering some specific diagnoses, yet still lacks detailed descriptions.

In summary, human annotation corroborates with our findings from Experiment 3. Given groundtruth conditions,
GPT-4V generated reports still lack comprehensive coverage of all relevant clinical findings and do not effectively summarize and organize medical conditions, compared with human-written reports.

% \newpage
\section{Limitations}
\label{sec:lim}

In this paper, we utilize the GPT-4 model series and Llama-3.2 to conduct a systematic evaluation of their capabilities in generating radiology reports. Comparisons with other general-domain multimodal LMMs, including Google's Gemini, are reserved for future research. 

Additionally, we use four common prompting strategies in our study and encourage future research to explore additional techniques, such as Self-Critique~\citep{shinn2023reflexion}, to verify the robustness of our findings. 
Due to resource constraints, we randomly select a subset of data for overall evaluation and use 50 samples for a human study. Besides, the human study is limited to two radiologists' subjective assessment, potentially influenced by their personal style and preferences.
While our human evaluation could be improved by recruiting more radiologists, we believe that GPT-4's poor performance may not justify a significantly larger human evaluation.
That said, our results suggest that finetuned open models may hold the potential of fitting into the current radiologist workflow if we can leverage medical image reasoning abilities of other models.

Despite these limitations, we believe the findings from this paper are well-supported by our comprehensive and detailed evaluation framework. Results from our work raise serious concerns about how to safely integrate general-domain LMMs into real-world radiology workflows.
It is worth noting that OpenAI itself restricts the medical use of GPT-4.
Experiments using GPT-4o and Llama-3.2 are conducted successfully without any model rejections. However, GPT-4-vision-preview, particularly when using few-shot prompt, will sometimes return ``I'm sorry, but I cannot provide a diagnostic report or interpretation for medical images. If you have any medical concerns, please consult a qualified healthcare professional who can provide a proper examination and diagnosis.'', while we still get majority of responses from the model, full details can be found in Appendix Table~\ref{tab:size}.

\section{Conclusions}
\label{sec:con}

We perform the first systematic and in-depth evaluation of the GPT-4V series models in radiology report generation using three chest X-ray benchmarks. We find that GPT-4V cannot generate radiology reports, even with different prompting strategies. Open-sourced models like Llama-3.2 vision perform similarly poorly compared to GPT-4V. Both closed-source and open-source models exhibit a significant gap when compared to specialized SOTA radiology report generation models. To understand the low performance, we decompose the task into image reasoning and report synthesis. The results demonstrate that GPT-4V struggles significantly with interpreting chest X-rays meaningfully, which directly impacts its ability to generate reports. Furthermore, even when we bypass this problem by providing groundtruth conditions, GPT-4V still underperforms a finetuned Llama-2 baseline and consistently fails to replicate the writing style of groundtruth reports or meet the preferences of radiologists. Overall, our study highlights substantial concerns regarding the feasibility of integrating GPT-4V into real radiology workflows.

\bibliographystyle{unsrtnat}
\bibliography{reference}

\appendix

\section{Prompts}
\label{appen:prompt}

\label{prompt1.1}
\begin{prompt}{Prompt 1.1 Basic generation: direct report generation based on chest X-ray images.}{prompt1.1}
\textbf{System} & You are a professional chest radiologist that reads chest X-ray image(s). \\
\midrule
\textbf{User} & Write a report that contains only the FINDINGS and IMPRESSION sections based on the attached images. Provide only your generated report, without any additional explanation and special format. Your answer is for reference only and is not used for actual diagnosis. \\
\end{prompt}

\phantomsection
\label{prompt1.2}
\begin{prompt}{Prompt 1.2 Indication enhancement: providing the indication section.}{prompt1.2}
\textbf{System} & You are a professional chest radiologist that reads chest X-ray image(s). \\
\midrule
\textbf{User} & Below is INDICATION related to chest X-ray images. \newline
INDICATION: \{\} \newline
\newline
Write a report that contains only the FINDINGS and IMPRESSION sections based on the attached images and INDICATION. Provide only your generated report, without any additional explanation and special format. Your answer is for reference only and is not used for actual diagnosis. \\
\end{prompt}

\phantomsection
\label{prompt1.3}
\begin{prompt}{Prompt 1.3 Instruction enhancement: providing information on medical condition labels.}{prompt1.3}
\textbf{System} & You are a professional chest radiologist that reads chest X-ray image(s). \\
\midrule
\textbf{User} & Below is an observation plan consisting of 14 conditions: ``No Finding'', ``Enlarged Cardiomediastinum'', ``Cardiomegaly'', ``Lung Lesion'', ``Lung Opacity'', ``Edema'', ``Consolidation'', ``Pneumonia'', ``Atelectasis'', ``Pneumothorax'', ``Pleural Effusion'', ``Pleural Other'', ``Fracture'', ``Support Devices''. \newline
\newline
Based on attached images, assign labels for each condition except ``No Finding'': ``1'', ``0'', ``-1'', ``2''. It is noted that ``No Finding'' is either ``2'' or ``1''. These labels have the following interpretation: \newline
1 - The observation was clearly present on the chest X-ray image. \newline
0 - The observation was absent on the chest X-ray image and was mentioned as negative. \newline
-1 - The observation was unclear if it exists. \newline
2 - The observation was absent but not explicitly mentioned. \newline
\newline
Based on labels you choose for each condition, write a report that contains only the FINDINGS and IMPRESSION sections. Don’t return any of your assigned labels. Provide only your generated report, without any additional explanation and special format. Your answer is for reference only and is not used for actual diagnosis.\\
\end{prompt}

\phantomsection
\label{prompt1.4}
\begin{prompt}{Prompt 1.4 Chain-of-Thought: step 1 - medical condition labeling; step 2 - report synthesis.}{prompt1.4}
\textbf{System} & You are a professional chest radiologist that reads chest X-ray image(s).\\
\midrule
\textbf{User} & Below is an observation plan consisting of 14 conditions: ``No Finding'', ``Enlarged Cardiomediastinum'', ``Cardiomegaly'', ``Lung Lesion'', ``Lung Opacity'', ``Edema'', ``Consolidation'', ``Pneumonia'', ``Atelectasis'', ``Pneumothorax'', ``Pleural Effusion'', ``Pleural Other'', ``Fracture'', ``Support Devices''. \newline
\newline
Based on attached images, assign labels for each condition except ``No Finding'': ``1'', ``0'', ``-1'', ``2''. It is noted that ``No Finding'' is either ``2'' or ``1''. These labels have the following interpretation: \newline
1 - The observation was clearly present on the chest X-ray image. \newline
0 - The observation was absent on the chest X-ray image and was mentioned as negative. \newline
-1 - The observation was unclear if it exists. \newline
2 - The observation was absent but not explicitly mentioned. \newline
\newline
The first step is to return one list of your assigned labels. For multiple images, assign the labels based on all images and return only one list of labels for the given 14 conditions. \newline
\newline
The second step is to write a report that contains only the FINDINGS and IMPRESSION sections based on labels you choose for each condition. \newline
\newline
Your answer is for reference only and is not used for actual diagnosis. Strictly follow the format below to provide your output.\newline
\newline
<LABEL> \newline
[ \newline
(``No Finding'', ``1''|``2''), \newline
(``Enlarged Cardiomediastinum'', ``0''|``1''|``2''|``-1''), \newline
(``Cardiomegaly'', ``0''|``1''|``2''|``-1''), \newline
(``Lung Lesion'', ``0''|``1''|``2''|``-1''), \newline
(``Lung Opacity'', ``0''|``1''|``2''|``-1''), \newline 
(``Edema'', ``0''|``1''|``2''|``-1''), \newline
(``Consolidation'', ``0''|``1''|``2''|``-1''), \newline
(``Pneumonia'', ``0''|``1''|``2''|``-1''), \newline
(``Atelectasis'', ``0''|``1''|``2''|``-1''), \newline
(``Pneumothorax'', ``0''|``1''|``2''|``-1''), \newline
(``Pleural Effusion'', ``0''|``1''|``2''|``-1''), \newline
(``Pleural Other'', ``0''|``1''|``2''|``-1''), \newline
(``Fracture'', ``0''|``1''|``2''|``-1''), \newline
(``Support Devices'', ``0''|``1''|``2''|``-1'') \newline
] \newline
</LABEL> \newline
<REPORT> \newline
FINDINGS: <findings> \newline
IMPRESSION: <impression> \newline
</REPORT> \\
\end{prompt}

\phantomsection
\label{prompt1.5-mimic}
\begin{prompt}{Prompt 1.5 Few-shot: few-shot in-context learning given a few examples (MIMIC).}{prompt1.5}
\textbf{System} & You are a professional chest radiologist that reads chest X-ray image(s).\\
\midrule
\textbf{User} & Write a report that contains only the FINDINGS and IMPRESSION sections based on the attached images. Provide only your generated report, without any additional explanation and special format. Your answer is for reference only and is not used for actual diagnosis. \newline
\newline
[.JPEG] \newline
FINDINGS: Single portable view of the chest is compared to previous exam from \_\_\_.  Enteric tube is seen with tip off the inferior field of view. Left PICC is seen; however, tip is not clearly delineated.  Persistent bibasilar effusions and a right pigtail catheter projecting over the lower chest.  There is possible right apical pneumothorax.  Superiorly, the lungs are clear of consolidation.  Cardiac silhouette is within normal limits. Osseous and soft tissue structures are unremarkable. \newline 
IMPRESSION: No significant interval change with bilateral \textbf{pleural effusions} with right pigtail catheter in the lower chest.  Possible small right apical pneumothorax. \newline
\newline
[.JPEG] \newline
FINDINGS: Frontal and lateral radiographs of the chest show hyperinflated lungs with flattened diaphragm, consistent with emphysema. Asymmetric opacity in the right middle lobe is concerning for pneumonia. No pleural effusion or pneumothorax is seen.  The cardiomediastinal contours are within normal limits aside from a tortuous aorta. \newline 
IMPRESSION: Right middle lobe opacity concerning for \textbf{pneumonia}. \newline
\newline
[.JPEG] \newline
FINDINGS: PA and lateral views of the chest provided. Midline sternotomy wires and mediastinal clips again noted. Suture is again noted in the right lower lung with adjacent rib resection. There is mild scarring in the right lower lung as on prior. There is no focal consolidation, large effusion or pneumothorax. No signs of congestion or edema. The heart remains moderately enlarged. The mediastinal contour is stable. \newline 
IMPRESSION: Postsurgical changes in the right hemithorax. Mild \textbf{cardiomegaly} unchanged. No edema or pneumonia. \newline
\newline
[.JPEG] \newline
FINDINGS: PA and lateral views of the chest provided. Biapical pleural parenchymal scarring noted.  No focal consolidation concerning for pneumonia. No effusion or pneumothorax.  No signs of congestion or edema.  Cardiomediastinal silhouette is stable with an unfolded thoracic aorta and top-normal heart size.  Bony structures are intact. \newline 
IMPRESSION: \textbf{No acute findings.} Top-normal heart size. \newline
\newline
[.JPEG] \newline\\
\end{prompt}

\phantomsection
\label{prompt1.5-openi}
\begin{prompt}{Prompt 1.5 Few-shot: few-shot in-context learning given a few examples (IU X-RAY).}{prompt1.5}
\textbf{System} & \multicolumn{1}{l}{You are a professional chest radiologist that reads chest X-ray image(s).} \\
\midrule
\textbf{User} & Write a report that contains only the FINDINGS and IMPRESSION sections based on the attached images. Provide only your generated report, without any additional explanation and special format. Your answer is for reference only and is not used for actual diagnosis. \newline
\newline
[.PNG] \newline
FINDINGS: 2 images. Heart size upper limits of normal. Mediastinal contours are maintained. The patient is mildly rotated. There is a small to moderate sized right apical pneumothorax which measures approximately 2.0 cm. No focal airspace consolidation is seen. Left chest is clear. No definite displaced bony injury is seen. Results called XXXX. XXXX XXXX p.m. XXXX, XXXX. \newline 
IMPRESSION: Small to moderate right apical \textbf{pneumothorax}. \newline
\newline
[.PNG] \newline
FINDINGS: The heart is normal in size and contour. There is focal airspace disease in the right middle lobe. There is no pneumothorax or effusion. \newline 
IMPRESSION: Focal airspace disease in the right middle lobe. This is most concerning for \textbf{pneumonia}. Recommend follow up to ensure resolution. \newline
\newline
[.PNG] \newline
FINDINGS: Stable cardiomegaly with vascular prominence without overt edema. No focal airspace disease. No large pleural effusion or pneumothorax. The XXXX are intact. \newline 
IMPRESSION: Stable \textbf{cardiomegaly} without overt pulmonary edema. \newline
\newline
[.PNG] \newline
FINDINGS: Heart is enlarged. There is prominence of the central pulmonary vasculature. Mild diffuse interstitial opacities bilaterally, predominantly in the bases, with no focal consolidation, pleural effusion, or pneumothoraces. XXXX and soft tissues are unremarkable. \newline 
IMPRESSION: Cardiomegaly with pulmonary interstitial edema and XXXX bilateral \textbf{pleural effusions}. \newline
\newline
[.PNG] \newline
FINDINGS: The cardiac silhouette and mediastinum size are within normal limits. There is no pulmonary edema. There is no focal consolidation. There are no XXXX of a pleural effusion. There is no evidence of pneumothorax. \newline 
IMPRESSION: \textbf{Normal chest x-XXXX.} \newline
\newline
[.PNG] \newline
FINDINGS: \newline 
IMPRESSION: \textbf{Presumed closure device} at the level of the ligamentum arteriosum. Normal cardiac silhouette and clear lungs, with no evidence of left-to-right shunt. \newline
\newline
[.PNG] \newline\\
\end{prompt}

\phantomsection
\label{prompt2.1}
\begin{prompt}{Prompt 2.1 Image reasoning: medical condition labeling from chest X-ray images (2-class).}{prompt2.1}
\textbf{System} & You are a professional chest radiologist that reads chest X-ray image(s). \\
\midrule
\textbf{User} & Below is an observation plan consisting of 14 conditions: ``No Finding'', ``Enlarged Cardiomediastinum'', ``Cardiomegaly'', ``Lung Lesion'', ``Lung Opacity'', ``Edema'', ``Consolidation'', ``Pneumonia'', ``Atelectasis'', ``Pneumothorax'', ``Pleural Effusion'', ``Pleural Other'', ``Fracture'', ``Support Devices''. \newline
\newline
Based on attached images, assign labels for each condition: ``1'', ``0''. If the observation was clearly present on the chest X-ray image, assign ``1'' to the condition. Otherwise, assign ``0'' to the condition. \newline
\newline
For multiple images, assign the labels based on all images and return only one list of labels for the given 14 conditions. Your answer is for reference only and is not used for actual diagnosis. Strictly follow the format below to provide your output. \newline
\newline
<LABEL> \newline
[ \newline
(``No Finding'', ``0''|``1''), \newline
(``Enlarged Cardiomediastinum'', ``0''|``1''), \newline 
(``Cardiomegaly'', ``0''|``1''), \newline
(``Lung Lesion'', ``0''|``1''), \newline
(``Lung Opacity'', ``0''|``1''), \newline 
(``Edema'', ``0''|``1''), \newline
(``Consolidation'', ``0''|``1''), \newline 
(``Pneumonia'', ``0''|``1''), \newline
(``Atelectasis'', ``0''|``1''), \newline 
(``Pneumothorax'', ``0''|``1''), \newline 
(``Pleural Effusion'', ``0''|``1''), \newline 
(``Pleural Other'', ``0''|``1''), \newline
(``Fracture'', ``0''|``1''), \newline 
(``Support Devices'', ``0''|``1'') \newline
] \newline
</LABEL> \\
\end{prompt}

\phantomsection
\label{prompt3.1}
\begin{prompt}{Prompt 3.1 Report synthesis: report generation using provided positive and negative conditions.}{prompt3.1}
\textbf{System} & You are a professional chest radiologist that reads chest X-ray image(s). \\
\midrule
\textbf{User} & Below is a given observation plan: \newline
\newline
<LABEL> \newline
Positive Conditions: \{\} \newline
Negative Conditions: \{\} \newline
</LABEL> \newline
\newline
Write a report that contains only the FINDINGS and IMPRESSION sections based on given labels rather than images. For positive conditions, you should clearly mention it in the report. For negative conditions, you should clearly mention in the report that there is no clear evidence of this condition. You should not mention any other conditions not listed above. Your answer is for reference only and is not used for actual diagnosis. Strictly follow the format below to provide your output. \newline
\newline 
<REPORT> \newline
FINDINGS: <findings> \newline
IMPRESSION: <impression> \newline
</REPORT> \\
\end{prompt}

\begin{prompt}{Prompt of finetuned LLaMA-2 report synthesis given groundtruth labels}{prompt-llama}
\textbf{System} & Write a radiology report that includes all given positive labels and negative labels. \\
\midrule
\textbf{User} & \textbf{Input:} \newline 
Positive labels: \{positive\_labels\} \newline 
Negative labels: \{negative\_labels\} \newline
\newline
\textbf{Output:} \{output\}\\
\end{prompt}

\section{Model Impementation Details}
\label{appen:model}

\paragraph{OpenAI API:}
We evaluate three datasets using Microsoft's Azure OpenAI service with GPT-4-vision-preview (11/15/2023) and GPT-4o.

\paragraph{Finetuning Llama-2:}
In the case of the MIMIC dataset, we selectively sample 10\% of the official training data, carefully ensuring there is no overlap with the 300-sample test set. For the IU X-ray dataset, we utilize the entire training set, which comprises 3,655 studies, and confirm that these too do not overlap with the test set. The fine-tuning process largely adheres to the default hyperparameters established by Stanford Alpaca~\citep{taori2023alpaca}. Our hardware includes four A100 GPUs, each equipped with 80GiB of memory, and operates on CUDA version 12.4. 

\paragraph{Llama-3.2-90B-Vision-Instruct:}
Llama-3.2-90B-Vision is the newest flagship open-source multimodal LLM. We choose it to provide a baseline reference. We access the model weights through huggingface and Meta. Our hardware for inference includes four A100 GPUs, each equipped with 80GiB of memory, and operates on CUDA version 12.4.

\paragraph{Code Availability:}
The source code for our project is publicly accessible on GitHub, enabling users and fellow researchers to review, utilize, or extend our implementations. You can find our repository at \url{https://github.com/ChicagoHAI/cxr-eval-gpt-4v.git}.

\section{Data}
\label{appen:data}

\paragraph{Data licenses:}
MIMIC-CXR license can be found at \url{https://physionet.org/content/mimic-cxr/view-license/2.0.0/}. 
IU X-RAY images are distributed under the terms of the Creative Commons Attribution-NonCommercial-NoDerivatives 4.0 International License (\url{http://creativecommons.org/licenses/by-nc-nd/4.0/}).

For MIMIC-CXR, we accessed the data by following the required steps on \url{https://physionet.org/content/mimic-cxr/2.0.0/}. We first registered and applied to be a credentialed user, and then completed the required training of CITI Data or Specimens Only Research. We also signed the data use agreement for the project before we get access to the dataset. We downloaded IU X-RAY dataset from \url{https://openi.nlm.nih.gov/faq}. Chexpert Plus dataset is available to download online and the license can be found at \url{https://stanfordaimi.azurewebsites.net/datasets/5158c524-d3ab-4e02-96e9-6ee9efc110a1}.

\paragraph{Preprocessing:}
To prepare the data for the OpenAI API, we first convert the DICOM images to JPEG format, which is required for compatibility with GPT-4V. Then we use base64 encoding to transform the binary image data into its corresponding UTF-8 string.

\paragraph{Ethical consideration of data:}
There is no substantial concerns around the data, since dataset are de-identified and do not contain harmful or offensive contents.

\section{Experiment Results}
\label{appen:result}

\subsection{GPT-4o Evaluation}
\label{appen:tabs}

\begin{figure*}
  \centering
  \includegraphics[scale=0.35]{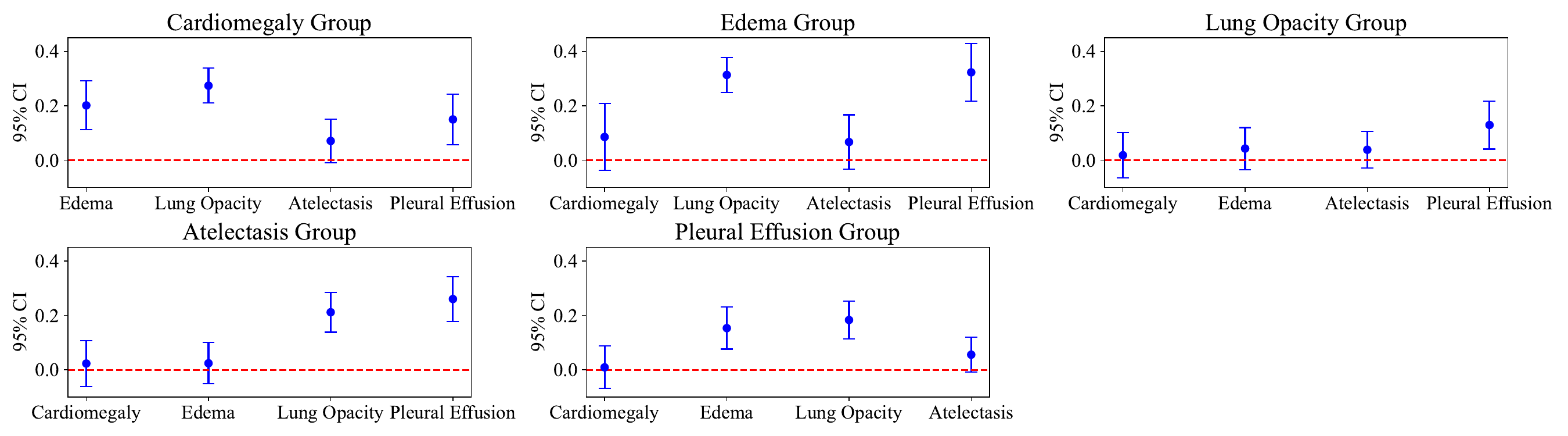}
  \caption{95\% Bootstrap confidence interval of top 5 conditions in our sample for GPT-4o.}
  \label{fig:ci-4o}
\end{figure*}

Overall performance of direct report generation across three datasets is shown in Table~\ref{tab:4o-openi}, Table~\ref{tab:4o-mimic} and Table~\ref{tab:4o-chexpert}.
Complete image reasoning results can be seen in Table~\ref{tab:gpt-4o-exp2-full}.
Complete Chi square test can be seen in Table~\ref{tab:gpt-4o-chi2-test}.
Bootstrap confidence intervals for most frequent conditions can be seen in Figure~\ref{fig:ci-4o}.

\subsection{GPT-4-vision-preview Evaluation}

\paragraph{Medical Restriction in API Usage}
Since OpenAI itself will restrict the medical use of GPT-4-vision-preview, the actual sample size of our expriments will be smaller than 300. Details can be checked in Table~\ref{tab:size}.

\paragraph{Overall Results}
Table \ref{tab:4v-mimic} and Table~\ref{tab:4v-openi} show all results on both findings section and impression section. 
Complete image reasoning results can be seen in Table~\ref{tab:4v-exp2}.
Complete Chi square test can be seen in Table~\ref{tab:4v-chi}.
Bootstrap confidence intervals for most frequent conditions can be seen in Figure~\ref{fig:ci-4v}.
Table~\ref{tab:4v-gt} show the generated results with groundtruth conditions.

\subsection{Llama-3.2-90B-Vision Evaluation}

Overall performance of direct report generation across three datasets is shown in Table~\ref{tab:llama-mimic}, Table~\ref{tab:llama-openi} and Table~\ref{tab:llama-chexpert}. It is noted that low hallucination of Llama-generated reports is probably due to its straightforward and assertive writing style, exemplified by sentences such as "The heart is enlarged," "There is a left pleural effusion," and "There is a right lung opacity."

\section{Hypothesis Test}
\label{appen:test}

\paragraph{Bootstrap Confidence Interval} 
We use \textbf{bootstrap confidence interval}~\citep{davison1997bootstrap} to test if GPT-4V labels one certain condition independently of the groundtruth condition group.
For this test, for each condition $i$ and group $j$, we define test statistic $\theta_{ij}$ as $P_{i}^{(j)}-P_{i}$ and null hypothesis $H_0$ as $\theta_{ij}=0$. We construct a 95\% confidence interval as $[\hat\theta_{ij,\; 0.025}^{(B)},\; \hat\theta_{ij,\; 0.975}^{(B)}]$ with 1000 bootstrap samples for each $\theta_{ij}$. Considering the sparsity of original study pool, we limit our choice of condition $i$ and group $j$ in six most frequent conditions in our subsample.

\paragraph{$\chi^2$ Test} Specifically, we use  a \textbf{$\chi^2$-test} to test if GPT-4V follows the same label distribution across different groups, i.e., testing the null hypothesis (\(H_0\)) that \(\mathbf{P}_{k} = \mathbf{P}_{k'} \) for any groups \(k\) and \(k'\). For the overall pool, we can construct a 13$\times$13 contingency table with each entry equal to $Y_{i}^{(j)}$ and then calculate expected count $E_{i}^{(j)}$ for each entry. Finally, report $\chi^2=\sum_i \sum_j \frac{(Y_{i}^{(j)}-E_{i}^{(j)})^2}{E_{i}^{(j)}}$. Considering the sparsity of original study pool, we report results of two different tables: (1) A modified table that replaces zero elements with 0.001; (2) A reduced table with only six most frequent conditions in subsample.

\begin{figure*}
  \centering
  \includegraphics[scale=0.55]{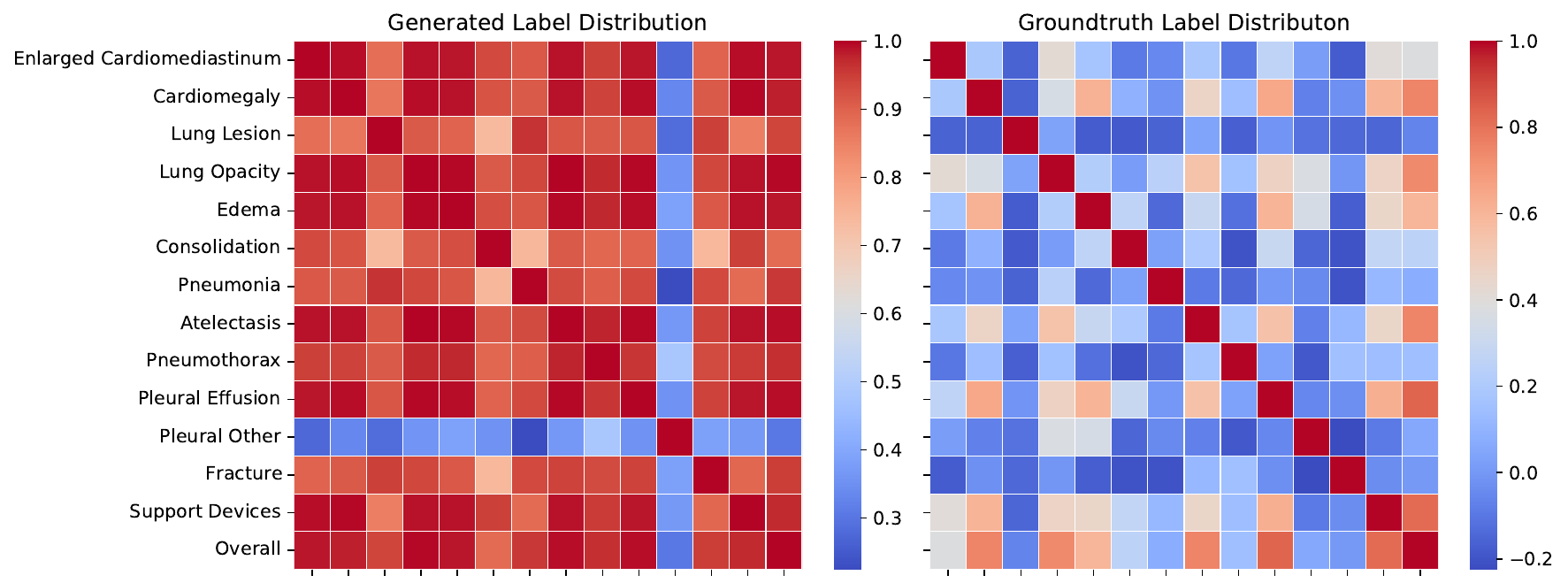}
  \caption{Correlations between distributions of Pr(Pos) for different condition groups (GPT-4-vision-preview).}
  \label{fig:corr-4v}
\end{figure*}

\begin{figure*}
  \centering
  \includegraphics[scale=0.55]{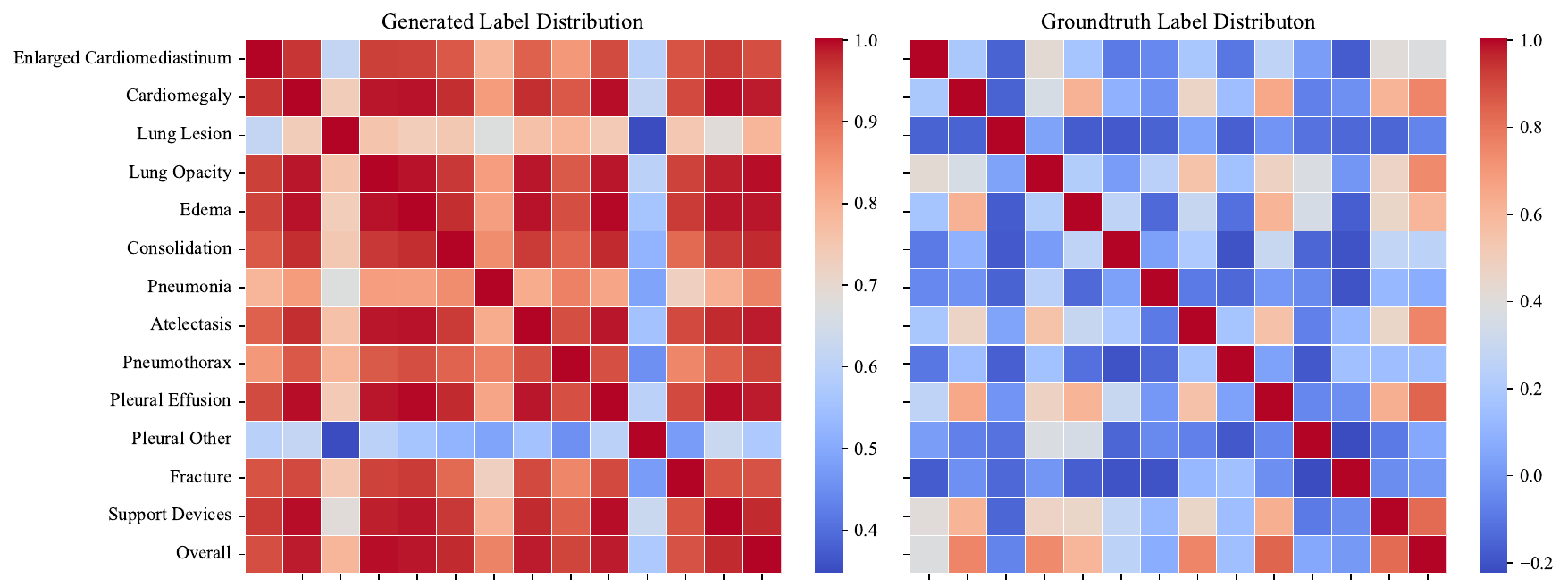}
  \caption{Correlations between distributions of Pr(Pos) for different condition groups (GPT-4o).}
  \label{fig:corr-4o}
\end{figure*}

\paragraph{Pearson Correlation Coefficient} We approximate \( P_{j}^{(k)} \) using \( \Pr(X_{ij}^{(k)} = 1) \) to obtain an estimator \( \mathbf{\widehat{P}_k} \) of \( \mathbf{P_k} \) for each group \( k \). Furthermore, we illustrate the correlation \( \text{Corr}(\mathbf{\widehat{P}_m}, \mathbf{\widehat{P}_n}) \) for all groups \( m \) and \( n \) in Figure~\ref{fig:corr-4v} and Figure~\ref{fig:corr-4o}. It is noted that the condition "Pleural Other" doesn't seem to be highly correlated with other groups. However, considering that "Pleural Other" only has one positive mention in groundtruth conditions and this can be treated as an outlier.

\begin{figure*}[htbp]
\centering
\includegraphics[scale=0.6]{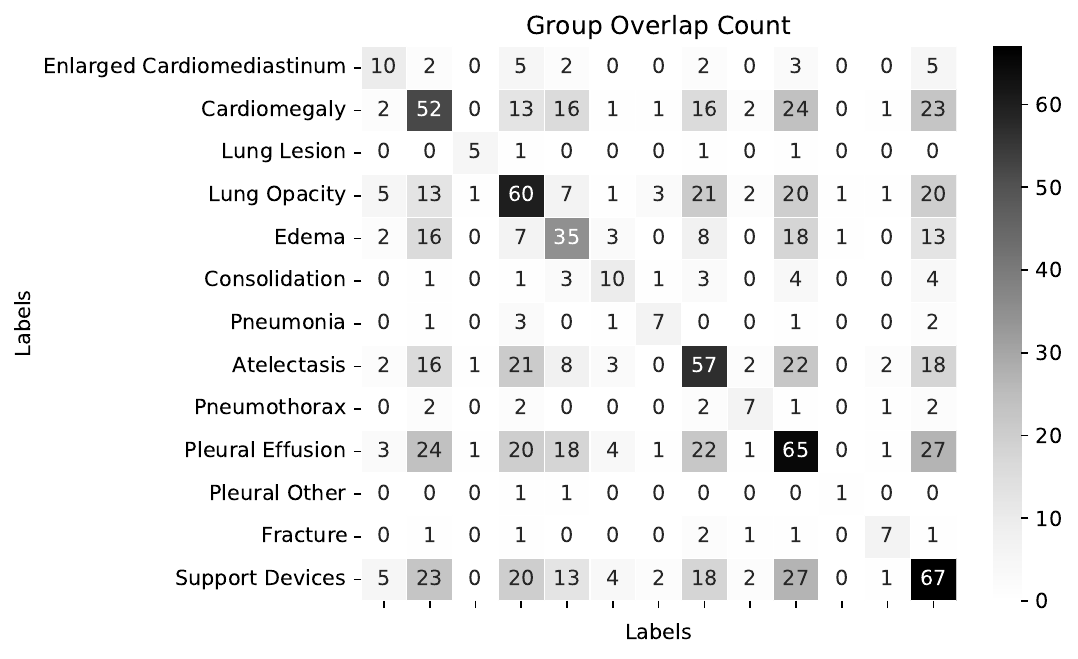}
\caption{Overlap count heatmap within groudtruth condition groups. The overlap between any two groups is relatively small compared to the original group sizes, which relate to the counts on the diagonal, ensuring that this overlap does not compromise the robustness of our results.}
\label{fig:overlap-heatmap}
\end{figure*}

\paragraph{Robustness Check}
We look into overlap issue between any two groups to further verify our results' robustness. We find out that it does not compromise the assumptions of the multinomial distribution or the robustness of Bootstrap CI results, but it could potentially inflate the Chi-square statistic, leading to incorrect conclusion about "no significance". To affirm the integrity of our Chi-square test results, we highlight two points: first, comparison with groundtruth studies in Table~\ref{tab:gpt-4o-exp2-chi} shows a distinct difference in p-values (small for groundtruth and large for generated studies), suggesting minimal impact of overlap on our conclusion that "GPT-4V generates reports following the same label distribution." Additionally, we also check the overlapped samples between groups and find them to be relatively small to original group sizes, with an average ratio of 13.1\%, as detailed in Figure~\ref{fig:overlap-heatmap}.

\section{Examples of Generated Reports}
\label{appen:reports}
The generated examples (GPT-4-vision-preview) by different prompting strategies are shown in table~\ref{tab:rad}.

\section{Human Reader Study}
\label{appen:human}
\begin{figure*}[h]
    \centering
    \includegraphics[height=0.8\textheight]{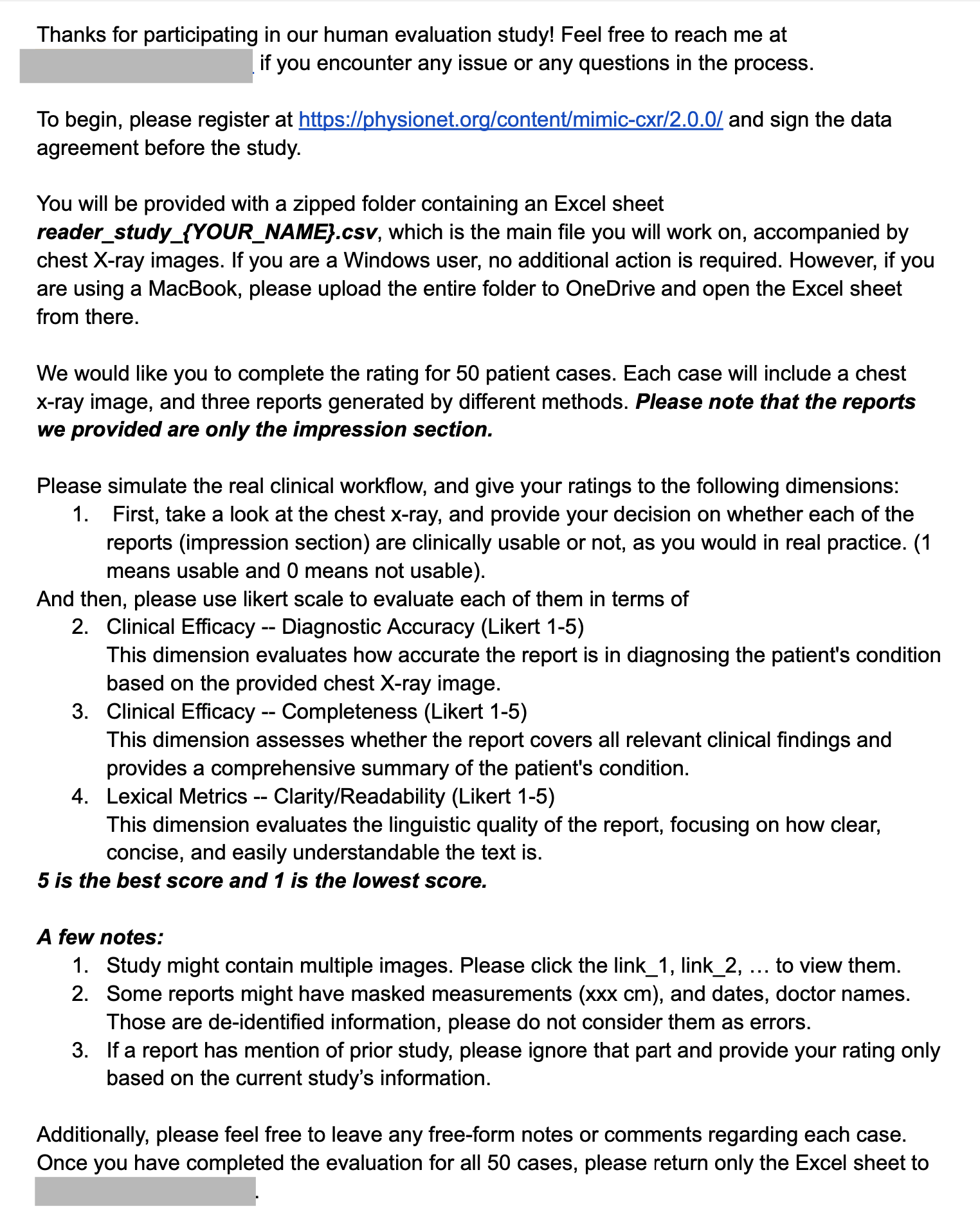}    
    \caption{Human reader study instructions.}
    \label{fig:human_instructions}
\end{figure*}

Instructions for human study are shown in Figure~\ref{fig:human_instructions}.

%-------------------------Tables and Figures-------------------------
\begin{table*}[htbp]
\centering
\small
\resizebox{\textwidth}{!}{
\begin{tabular}{@{}l@{\hspace{2pt}}l@{\hspace{2pt}}l@{\hspace{2pt}}l@{\hspace{2pt}}l@{\hspace{2pt}}l@{\hspace{2pt}}l@{\hspace{2pt}}l@{\hspace{2pt}}l@{\hspace{2pt}}l@{\hspace{2pt}}l@{\hspace{2pt}}}
\toprule
\textbf{Experiment} & \multicolumn{4}{c}{\textbf{Lexical metrics}} & \multicolumn{6}{c}{\textbf{Clinical Efficacy Metrics}} \\
\cmidrule(lr){2-5} \cmidrule(l){6-11}
& \textbf{BLEU-1} & \textbf{BLEU-4} & \textbf{ROUGE-L} & \textbf{METEOR} & \textbf{Pos F1} & \textbf{Pos F1@5} & \textbf{RadGraph} & \textbf{Neg F1} & \textbf{Neg F1@5} & \textbf{Hall.$\downarrow$} \\
\midrule
\multicolumn{11}{c}{\textsc{IU X-ray (Findings)}} \\
\midrule
1.1 & 0.316 & 0.045 & 0.238 & 0.311 & 0.059 & 0.045 & 0.203 & 0.272 & 0.397 & 0.354 \\
1.2 & 0.330 & 0.049 & 0.242 & 0.323 & 0.077 & 0.098 & 0.214 & 0.285 & 0.419 & 0.269 \\
1.3 & 0.238 & 0.030 & 0.207 & 0.283 & 0.081 & 0.146 & 0.174 & 0.298 & 0.458 & 0.169 \\
1.4 & 0.239 & 0.024 & 0.194 & 0.231 & 0.077 & 0.161 & 0.144 & 0.285 & 0.440 & 0.138 \\
1.5 & 0.279 & 0.044 & 0.243 & 0.250 & 0.037 & 0.031 & 0.187 & 0.240 & 0.379 & 0.097 \\
\midrule
\multicolumn{11}{c}{\textsc{IU X-ray (Impression)}} \\
\midrule
1.1 & 0.212 & 0.012 & 0.194 & 0.269 & 0.087 & 0.105 & 0.104 & 0.000 & 0.000 & 0.303 \\
1.2 & 0.248 & 0.014 & 0.251 & 0.341 & 0.173 & 0.135 & 0.153 & 0.071 & 0.051 & 0.307 \\
1.3 & 0.209 & 0.012 & 0.205 & 0.281 & 0.111 & 0.206 & 0.106 & 0.000 & 0.000 & 0.177 \\
1.4 & 0.181 & 0.007 & 0.143 & 0.177 & 0.115 & 0.255 & 0.060 & 0.000 & 0.000 & 0.197 \\
1.5 & 0.138 & 0.001 & 0.077 & 0.124 & 0.067 & 0.107 & 0.034 & 0.010 & 0.025 & 0.211 \\
\bottomrule
\end{tabular}
}
\caption{Direct report generation performance comparison for IU X-ray findings and impressions (GPT-4o).}
\label{tab:4o-openi}
\end{table*}

\begin{table*}[htbp]
\centering
\resizebox{\textwidth}{!}{
\begin{tabular}{@{}l@{\hspace{2pt}}l@{\hspace{2pt}}l@{\hspace{2pt}}l@{\hspace{2pt}}l@{\hspace{2pt}}l@{\hspace{2pt}}l@{\hspace{2pt}}l@{\hspace{2pt}}l@{\hspace{2pt}}l@{\hspace{2pt}}l@{\hspace{2pt}}}
\toprule
\textbf{Experiment} & \multicolumn{4}{c}{\textbf{Lexical metrics}} & \multicolumn{6}{c}{\textbf{Clinical Efficacy Metrics}} \\
\cmidrule(lr){2-5} \cmidrule(l){6-11}
& \textbf{BLEU-1} & \textbf{BLEU-4} & \textbf{ROUGE-L} & \textbf{METEOR} & \textbf{Pos F1} & \textbf{Pos F1@5} & \textbf{RadGraph} & \textbf{Neg F1} & \textbf{Neg F1@5} & \textbf{Hall.$\downarrow$} \\
\midrule
\multicolumn{11}{c}{\textsc{MIMIC-CXR (Findings)}} \\
\midrule
1.1 & 0.331 & 0.031 & 0.225 & 0.282 & 0.134 & 0.153 & 0.163 & 0.244 & 0.418 & 0.421 \\
1.2 & 0.333 & 0.044 & 0.234 & 0.290 & 0.273 & 0.255 & 0.182 & 0.259 & 0.431 & 0.383 \\
1.3 & 0.276 & 0.028 & 0.203 & 0.264 & 0.155 & 0.230 & 0.154 & 0.290 & 0.503 & 0.298 \\
1.4 & 0.242 & 0.017 & 0.186 & 0.201 & 0.172 & 0.272 & 0.119 & 0.249 & 0.443 & 0.180 \\
1.5 & 0.337 & 0.055 & 0.257 & 0.301 & 0.170 & 0.203 & 0.188 & 0.279 & 0.481 & 0.934 \\
\midrule
\multicolumn{11}{c}{\textsc{MIMIC-CXR (Impression)}} \\
\midrule
1.1 & 0.159 & 0.006 & 0.142 & 0.185 & 0.123 & 0.152 & 0.077 & 0.038 & 0.081 & 0.587 \\
1.2 & 0.203 & 0.014 & 0.195 & 0.238 & 0.277 & 0.258 & 0.124 & 0.048 & 0.126 & 0.547 \\
1.3 & 0.168 & 0.009 & 0.162 & 0.200 & 0.156 & 0.243 & 0.095 & 0.042 & 0.101 & 0.359 \\
1.4 & 0.120 & 0.004 & 0.112 & 0.130 & 0.177 & 0.288 & 0.055 & 0.008 & 0.020 & 0.397 \\
1.5 & 0.155 & 0.007 & 0.153 & 0.186 & 0.144 & 0.232 & 0.073 & 0.035 & 0.091 & 0.477 \\
\bottomrule
\end{tabular}
}
\caption{Performance comparison for MIMIC-CXR findings and impressions (GPT-4o).}
\label{tab:4o-mimic}
\end{table*}

\begin{table*}[htbp]
\centering
\resizebox{\textwidth}{!}{
\begin{tabular}{@{}l@{\hspace{2pt}}l@{\hspace{2pt}}l@{\hspace{2pt}}l@{\hspace{2pt}}l@{\hspace{2pt}}l@{\hspace{2pt}}l@{\hspace{2pt}}l@{\hspace{2pt}}l@{\hspace{2pt}}l@{\hspace{2pt}}l@{\hspace{2pt}}}
\toprule
\textbf{Experiment} & \multicolumn{4}{c}{\textbf{Lexical metrics}} & \multicolumn{6}{c}{\textbf{Clinical Efficacy Metrics}} \\
\cmidrule(lr){2-5} \cmidrule(l){6-11}
& \textbf{BLEU-1} & \textbf{BLEU-4} & \textbf{ROUGE-L} & \textbf{METEOR} & \textbf{Pos F1} & \textbf{Pos F1@5} & \textbf{RadGraph} & \textbf{Neg F1} & \textbf{Neg F1@5} & \textbf{Hall.$\downarrow$} \\
\midrule
\multicolumn{11}{c}{\textsc{ChexPert Plus (Findings)}} \\
\midrule
1.1 & 0.237 & 0.015 & 0.176 & 0.191 & 0.228 & 0.191 & 0.112 & 0.118 & 0.174 & 0.516 \\
1.3 & 0.191 & 0.007 & 0.159 & 0.172 & 0.210 & 0.325 & 0.101 & 0.158 & 0.253 & 0.371 \\
1.4 & 0.166 & 0.011 & 0.155 & 0.139 & 0.234 & 0.339 & 0.077 & 0.160 & 0.262 & 0.339 \\
1.5 & 0.171 & 0.007 & 0.158 & 0.149 & 0.188 & 0.224 & 0.094 & 0.104 & 0.135 & 0.411 \\
3.1 & 0.085 & 0.003 & 0.150 & 0.108 & 0.483 & 0.724 & 0.080 & 0.148 & 0.250 & 0.226 \\
\midrule
\multicolumn{11}{c}{\textsc{ChexPert Plus (Impression)}} \\
\midrule
1.1 & 0.105 & 0.001 & 0.106 & 0.090 & 0.163 & 0.198 & 0.008 & 0.013 & 0.035 & 0.680 \\
1.3 & 0.109 & 0.002 & 0.105 & 0.095 & 0.172 & 0.261 & 0.008 & 0.042 & 0.085 & 0.377 \\
1.4 & 0.066 & 0.000 & 0.093 & 0.069 & 0.192 & 0.242 & 0.012 & 0.000 & 0.000 & 0.400 \\
1.5 & 0.084 & 0.002 & 0.131 & 0.088 & 0.134 & 0.161 & 0.029 & 0.043 & 0.111 & 0.370 \\
3.1 & 0.106 & 0.003 & 0.164 & 0.116 & 0.854 & 0.947 & 0.022 & 0.662 & 0.924 & 0.480 \\
\bottomrule
\end{tabular}
}
\caption{Performance comparison for ChexPert Plus findings and impressions (GPT-4o).}
\label{tab:4o-chexpert}
\end{table*}

\begin{table*}[t]
\centering
\resizebox{.9\textwidth}{!}{
\begin{tabular}{lccccc}
\toprule
\multirow{2}{*}{\textbf{Metric}} & \multicolumn{2}{c}{\textbf{MIMIC-CXR}} & \multicolumn{2}{c}{\textbf{CheXpert Plus}} \\
\cmidrule(lr){2-3} \cmidrule(lr){4-5}
& \textbf{Chain-of-Thought (1st Step)} & \textbf{Image Reasoning} & \textbf{Chain-of-Thought (1st Step)} & \textbf{Image Reasoning} \\
\midrule
\textbf{Positive F1} & 0.195 & 0.161 & 0.221 & 0.189 \\
\textbf{Positive F1@5} & 0.298 & 0.242 & 0.257 & 0.247 \\
\bottomrule
\end{tabular}
}
\caption{
Image reasoning performance of GPT-4o on chest X-ray images. 
}
\label{tab:gpt-4o-exp2-full}
\end{table*}

\begin{table*}[h]
\centering
\resizebox{\textwidth}{!}{
\begin{tabular}{lcccccccc}
\toprule
\multirow{2}{*}{\textbf{}} & \multicolumn{4}{c}{\textbf{MIMIC}} & \multicolumn{4}{c}{\textbf{CheXpert Plus}} \\
\cmidrule(lr){2-5} \cmidrule(lr){6-9}
& \textbf{Overall (GT)} & \textbf{Overall (Gen)} & \textbf{Top 6 (GT)} & \textbf{Top 6 (Gen)} & \textbf{Overall (GT)} & \textbf{Overall (Gen)} & \textbf{Top 6 (GT)} & \textbf{Top 6 (Gen)} \\
\midrule
$\chi^2$ statistic & 1770.38 & 66.05 & 317.86 & 8.71 & 1546.86 & 53.24 & 309.31 & 3.88 \\
p-value & p<1e-1 & 1 & p<1e-1 & 0.9989 & p<1e-1 & 1 & p<1e-1 & 1 \\
df. & 144 & 144 & 25 & 25 & 144 & 144 & 25 & 25 \\
\bottomrule
\end{tabular}
}
  \caption{$\chi^2$-test for homogeneity of label distribution across different condition groups (GPT-4o).}
  \label{tab:gpt-4o-chi2-test}
\end{table*}

\begin{table*}[t]
\centering
\small
\begin{threeparttable}
\begin{tabular}{@{}p{2cm}@{\hspace{8pt}}c@{\hspace{2pt}}c@{\hspace{2pt}}c@{\hspace{2pt}}c@{\hspace{2pt}}c@{\hspace{2pt}}c@{\hspace{2pt}}c@{}}
\toprule
\multirow{2}{*}{\textbf{Experiment}} & \multicolumn{3}{c}{\textbf{IU X-RAY}} & \multicolumn{3}{c}{\textbf{MIMIC-CXR}} \\
\cmidrule(r){2-4} \cmidrule(r){5-7}
& \textbf{IMPRESSION} & \textbf{FINDINGS} & \textbf{Labels} & \textbf{IMPRESSION} & \textbf{FINDINGS} & \textbf{Labels} \\
\midrule
1.1 & 298/300 & 259/260 & - & 300/300 & 183/183 & - \\
1.2 & 295/300 & 259/260 & - & 300/300 & 183/183 & - \\
1.3 & 278/300 & 241/260 & - & 300/300 & 183/183 & - \\
1.4 & 258/300 & 223/260 & - & 300/300 & 183/183 & - \\
1.5 & 118/300 & 101/260 & - & 83/300 & 61/183 & - \\
\midrule
2.1 & - & - & 237/300 & - & - & 300/300 \\
\midrule
3.1 & 293/300 & 253/260 & - & 297/300 & 182/183 & - \\
\bottomrule
\end{tabular}
\caption{Summary of actual sample size for GPT-4-vision-preview across different experiments.}
\label{tab:size}
\end{threeparttable}
\end{table*}

\begin{table*}[htbp]
\centering
\small
\resizebox{\textwidth}{!}{
\begin{tabular}{@{}p{2cm}@{\hspace{8pt}}l@{\hspace{2pt}}l@{\hspace{2pt}}l@{\hspace{2pt}}l@{\hspace{2pt}}l@{\hspace{2pt}}l@{\hspace{2pt}}l@{\hspace{2pt}}l@{\hspace{2pt}}l@{\hspace{2pt}}l@{}}
\toprule
\multirow{2}{*}{\textbf{Experiment}} & \multicolumn{4}{c}{\textbf{Lexical metrics}} & \multicolumn{6}{c}{\textbf{Clinic Efficacy Metrics}} \\
\cmidrule(lr){2-5} \cmidrule(l){6-11}
& \textbf{BLEU-1} & \textbf{BLEU-4} & \textbf{ROUGE} & \textbf{METEOR} & \textbf{Pos F1} & \textbf{Pos F1@5} & \textbf{Rad. F1} & \textbf{Neg F1} & \textbf{Neg F1@5} & \textbf{Hall.$\downarrow$} \\
\midrule
\multicolumn{11}{c}{\textsc{MIMIC-CXR (Findings)}} \\
\midrule
1.1 & 0.299 & 0.035 & 0.214 & 0.279 & 0.117 & 0.124 & 0.135 & 0.241 & 0.396 & 0.563 \\
1.2 & 0.323 & 0.042 & 0.227 & 0.294 & 0.181 & 0.194 & 0.159 & 0.272 & 0.464 & 0.410 \\
1.3 & 0.265 & 0.019 & 0.186 & 0.262 & 0.134 & 0.236 & 0.109 & 0.237 & 0.437 & 0.607 \\
1.4 & 0.236 & 0.008 & 0.176 & 0.202 & 0.151 & 0.233 & 0.080 & 0.151 & 0.328 & 0.563 \\
1.5 & 0.294 & 0.053 & 0.223 & 0.293 & 0.085 & 0.036 & 0.149 & 0.251 & 0.462 & 1.000\\
\midrule
\multicolumn{11}{c}{\textsc{MIMIC-CXR (Impression)}} \\
\midrule
1.1 & 0.135 & 0.018 & 0.119 & 0.161 & 0.118 & 0.160 & 0.071 & 0.004 & 0.001 & 0.687 \\
1.2 & 0.176 & 0.021 & 0.163 & 0.200 & 0.185 & 0.200 & 0.101 & 0.037 & 0.096 & 0.610 \\
1.3 & 0.141 & 0.009 & 0.120 & 0.174 & 0.141 & 0.228 & 0.068 & 0.026 & 0.067 & 0.593 \\
1.4 & 0.113 & 0.002 & 0.107 & 0.133 & 0.150 & 0.255 & 0.058 & 0.023 & 0.061 & 0.607 \\
1.5 & 0.163 & 0.011 & 0.160 & 0.242 & 0.070 & 0.072 & 0.088 & 0.000 & 0.000 & 0.578 \\
\bottomrule
\end{tabular}
}
\caption{Direct report generation performance 
comparison for MIMIC-CXR findings and impressions (GPT-4-vision-preview).}
\label{tab:4v-mimic}
\end{table*}

\begin{table*}[htbp]
\centering
\resizebox{\textwidth}{!}{
\begin{tabular}{@{}p{2cm}@{\hspace{8pt}}l@{\hspace{2pt}}l@{\hspace{2pt}}l@{\hspace{2pt}}l@{\hspace{2pt}}l@{\hspace{2pt}}l@{\hspace{2pt}}l@{\hspace{2pt}}l@{\hspace{2pt}}l@{\hspace{2pt}}l@{}}
\toprule
\multirow{2}{*}{\textbf{Experiment}} & \multicolumn{4}{c}{\textbf{Lexical metrics}} & \multicolumn{6}{c}{\textbf{Clinic Efficacy Metrics}} \\
\cmidrule(lr){2-5} \cmidrule(l){6-11}
& \textbf{BLEU-1} & \textbf{BLEU-4} & \textbf{ROUGE} & \textbf{METEOR} & \textbf{Pos F1} & \textbf{Pos F1@5} & \textbf{Rad. F1} & \textbf{Neg F1} & \textbf{Neg F1@5} & \textbf{Hall.$\downarrow$} \\
\midrule
\multicolumn{11}{c}{\textsc{IU X-Ray (findings)}} \\
\midrule
1.1 & 0.278 & 0.038 & 0.218 & 0.326 & 0.030 & 0.024 & 0.178 & 0.284 & 0.429 & 0.494 \\
1.2 & 0.282 & 0.042 & 0.216 & 0.328 & 0.023 & 0.010 & 0.174 & 0.308 & 0.475 & 0.614 \\
1.3 & 0.237 & 0.027 & 0.189 & 0.281 & 0.053 & 0.052 & 0.140 & 0.265 & 0.429 & 0.523 \\
1.4 & 0.233 & 0.016 & 0.179 & 0.235 & 0.072 & 0.119 & 0.105 & 0.226 & 0.402 & 0.619 \\
1.5 & 0.325 & 0.037 & 0.247 & 0.318 & 0.061 & 0.080 & 0.191 & 0.290 & 0.455 & 0.287 \\
\midrule
\multicolumn{11}{c}{\textsc{IU X-Ray (impression)}} \\
\midrule
1.1 & 0.219 & 0.019 & 0.232 & 0.295 & 0.036 & 0.041 & 0.155 & 0.000 & 0.000 & 0.275 \\
1.2 & 0.209 & 0.021 & 0.215 & 0.295 & 0.058 & 0.060 & 0.169 & 0.020 & 0.052 & 0.410 \\
1.3 & 0.202 & 0.013 & 0.205 & 0.287 & 0.041 & 0.051 & 0.142 & 0.041 & 0.106 & 0.435 \\
1.4 & 0.172 & 0.009 & 0.155 & 0.216 & 0.052 & 0.100 & 0.097 & 0.000 & 0.000 & 0.628 \\
1.5 & 0.247 & 0.026 & 0.243 & 0.279 & 0.036 & 0.047 & 0.138 & 0.026 & 0.067 & 0.263 \\
\bottomrule
\end{tabular}
}
\caption{Direct report generation performance comparison for IU X-RAY findings and impressions (GPT-4-vision-preview).}
\label{tab:4v-openi}
\end{table*}

\begin{table*}[t]
\centering
\resizebox{\textwidth}{!}{
\begin{tabular}{lccccc}
\toprule
\multirow{2}{*}{\textbf{Metric}} & \multicolumn{2}{c}{\textbf{MIMIC-CXR}} & \multicolumn{2}{c}{\textbf{IU X-RAY}} \\
\cmidrule(lr){2-3} \cmidrule(lr){4-5}
& \textbf{Chain-of-Thought (1st Step)} & \textbf{Image Reasoning} & \textbf{Chain-of-Thought (1st Step)} & \textbf{Image Reasoning} \\
\midrule
\textbf{Positive F1} & 0.166 & 0.146 & 0.072 & 0.049 \\
\textbf{Positive F1@5} & 0.261 & 0.208 & 0.095 & 0.056 \\
\bottomrule
\end{tabular}
}
\caption{
Image reasoning performance of GPT-4-vision-preview on chest X-ray images. 
}
\label{tab:4v-exp2}
\end{table*}

\begin{table*}[t]
\centering
  \small
  % \resizebox{\columnwidth}{!}{
  \begin{tabular}{ccccc}
    \toprule
    \multirow{2}{*}{\bf Statistics} & \multicolumn{2}{c}{\bf Overall} & \multicolumn{2}{c}{\bf Top 6 Conditions} \\
    \cmidrule(r){2-3}
    \cmidrule(l){4-5}
    & \bf Groundtruth & \bf GPT-4V & \bf Groundtruth & \bf GPT-4V \\
    \midrule
    $\chi^2$ statistic & 1770.38 & 74.25 & 317.86 & 6.11 \\
  p-value & p < 1e-4 & 1.00 & p < 1e-4 & 1.00 \\
    df. & 144 & 144 & 25 & 25 \\
  \bottomrule
\end{tabular}
% }
  \caption{$\chi^2$-test for homogeneity of label distribution across different condition groups (GPT-4-vision-preview).}
    \label{tab:4v-chi}
\end{table*}

\begin{table*}[t]
\centering
\resizebox{\textwidth}{!}{
\begin{threeparttable}
\begin{tabular}{lcccccccccc}
\toprule
\multirow{2}{*}{\textbf{Experiment}} & \multicolumn{4}{c}{\textbf{Lexical metrics}} & \multicolumn{6}{c}{\textbf{Clinic Efficacy Metrics}} \\
\cmidrule(lr){2-5} \cmidrule(l){6-11}
& \textbf{BLEU-1} & \textbf{BLEU-4} & \textbf{ROUGE} & \textbf{METEOR} & \textbf{Pos F1} & \textbf{Pos F1@5} & \textbf{Rad. F1} & \textbf{Neg F1} & \textbf{Neg F1@5} & \textbf{Hall.}$\downarrow$ \\
\midrule
\multicolumn{11}{c}{\textsc{MIMIC-CXR}} \\
GPT-4V & 0.135 & 0.018 & 0.119 & 0.161 & 0.118 & 0.160 & 0.071 & 0.004 & 0.001 & 0.687 \\
GPT-4V (gt) & 0.176 & 0.007 & 0.185 & 0.179 & 0.885 & \bf 0.977 & 0.103 & 0.584 & \bf 0.958& \bf 0.431\\
 LLaMA-2 (gt) & \bf 0.301 & \bf 0.094 & \bf 0.330& \bf 0.348& \bf 0.923 & 0.957& \bf 0.286& \bf 0.703&  0.941&  0.710\\
\midrule
\multicolumn{11}{c}{\textsc{IU X-Ray}} \\
GPT-4V & 0.219 & 0.019 & 0.232 & 0.295 & 0.036 & 0.041 & 0.155 & 0.000 & 0.000 & 0.275 \\
GPT-4V (gt) & 0.216 & 0.003 & 0.229 & 0.207 & 0.852 & 0.919 & 0.089 & \bf 0.630 & 0.868 & 0.235 \\
 LLaMA-2 (gt) & \bf 0.454 & \bf 0.124 & \bf 0.460 & \bf 0.441 & \bf 0.871 & \bf 0.928 & \bf 0.297 & 0.627 & \bf 0.963 & \bf 0.110 \\
\bottomrule
\end{tabular}
\begin{tablenotes}
\item All metrics are evaluated on the impression section. 
\end{tablenotes}
\end{threeparttable}
}
\caption{Performance in report generation with groundtruth conditions (GPT-4-vision-preiew).
}
\label{tab:4v-gt}
\end{table*}

\begin{table*}[htbp]
\centering
\resizebox{\textwidth}{!}{
\begin{tabular}{@{}l@{\hspace{2pt}}l@{\hspace{2pt}}l@{\hspace{2pt}}l@{\hspace{2pt}}l@{\hspace{2pt}}l@{\hspace{2pt}}l@{\hspace{2pt}}l@{\hspace{2pt}}l@{\hspace{2pt}}l@{\hspace{2pt}}l@{\hspace{2pt}}}
\toprule
\textbf{Experiment} & \multicolumn{4}{c}{\textbf{Lexical metrics}} & \multicolumn{6}{c}{\textbf{Clinical Efficacy Metrics}} \\
\cmidrule(lr){2-5} \cmidrule(l){6-11}
& \textbf{BLEU-1} & \textbf{BLEU-4} & \textbf{ROUGE-L} & \textbf{METEOR} & \textbf{Pos F1} & \textbf{Pos F1@5} & \textbf{RadGraph} & \textbf{Neg F1} & \textbf{Neg F1@5} & \textbf{Hall.} \\
\midrule
\multicolumn{10}{c}{\textsc{MIMIC-CXR (Findings)}} \\
\midrule
1.1 & 0.258 & 0.026 & 0.236 & 0.216 & 0.159 & 0.266 & 0.123 & 0.200 & 0.267 & 0.607 \\
1.2 & 0.251 & 0.020 & 0.249 & 0.215 & 0.123 & 0.217 & 0.144 & 0.214 & 0.305 & 0.710 \\
1.3 & 0.110 & 0.012 & 0.114 & 0.122 & 0.118 & 0.221 & 0.073 & 0.054 & 0.111 & 0.262 \\
1.4 & 0.015 & 0.000 & 0.051 & 0.032 & 0.172 & 0.300 & 0.010 & 0.024 & 0.045 & 0.197 \\
1.5 & 0.233 & 0.023 & 0.233 & 0.202 & 0.144 & 0.249 & 0.120 & 0.197 & 0.270 & 0.579 \\
\midrule
\multicolumn{10}{c}{\textsc{MIMIC-CXR (Impression)}} \\
\midrule
1.1 & 0.089 & 0.001 & 0.085 & 0.094 & 0.121 & 0.231 & 0.036 & 0.000 & 0.000 & 0.161 \\
1.2 & 0.085 & 0.000 & 0.076 & 0.088 & 0.132 & 0.186 & 0.028 & 0.000 & 0.000 & 0.200 \\
1.3 & 0.089 & 0.001 & 0.089 & 0.100 & 0.141 & 0.215 & 0.046 & 0.045 & 0.117 & 0.049 \\
1.4 & 0.095 & 0.001 & 0.068 & 0.081 & 0.194 & 0.334 & 0.022 & 0.000 & 0.000 & 0.990 \\
1.5 & 0.086 & 0.000 & 0.081 & 0.088 & 0.111 & 0.216 & 0.035 & 0.000 & 0.000 & 0.160 \\
\bottomrule
\end{tabular}
}
\caption{Direct report generation performance comparison for MIMIC-CXR findings and impressions (Llama-3.2-90B-Vision-Instruct).}
\label{tab:llama-mimic}
\end{table*}

\begin{table*}[htbp]
\centering
\resizebox{\textwidth}{!}{
\begin{tabular}{@{}l@{\hspace{2pt}}l@{\hspace{2pt}}l@{\hspace{2pt}}l@{\hspace{2pt}}l@{\hspace{2pt}}l@{\hspace{2pt}}l@{\hspace{2pt}}l@{\hspace{2pt}}l@{\hspace{2pt}}l@{\hspace{2pt}}l@{\hspace{2pt}}}
\toprule
\textbf{Experiment} & \multicolumn{4}{c}{\textbf{Lexical metrics}} & \multicolumn{6}{c}{\textbf{Clinical Efficacy Metrics}} \\
\cmidrule(lr){2-5} \cmidrule(l){6-11}
& \textbf{BLEU-1} & \textbf{BLEU-4} & \textbf{ROUGE-L} & \textbf{METEOR} & \textbf{Pos F1} & \textbf{Pos F1@5} & \textbf{RadGraph} & \textbf{Neg F1} & \textbf{Neg F1@5} & \textbf{Hall.} \\
\midrule
\multicolumn{11}{c}{\textsc{IU X-ray (Findings)}} \\
\midrule
1.1 & 0.248 & 0.018 & 0.231 & 0.231 & 0.064 & 0.075 & 0.146 & 0.220 & 0.255 & 0.404 \\
1.2 & 0.244 & 0.019 & 0.239 & 0.216 & 0.021 & 0.056 & 0.156 & 0.221 & 0.265 & 0.342 \\
1.3 & 0.203 & 0.027 & 0.190 & 0.187 & 0.068 & 0.082 & 0.138 & 0.170 & 0.308 & 0.142 \\
1.4 & 0.015 & 0.001 & 0.053 & 0.037 & 0.104 & 0.214 & 0.008 & 0.010 & 0.012 & 0.677 \\
1.5 & 0.231 & 0.018 & 0.236 & 0.221 & 0.061 & 0.094 & 0.158 & 0.227 & 0.278 & 0.579 \\
\midrule
\multicolumn{11}{c}{\textsc{IU X-ray (Impression)}} \\
\midrule
1.1 & 0.117 & 0.000 & 0.063 & 0.102 & 0.054 & 0.087 & 0.019 & 0.000 & 0.000 & 0.057 \\
1.2 & 0.117 & 0.000 & 0.058 & 0.094 & 0.080 & 0.197 & 0.015 & 0.000 & 0.000 & 0.043 \\
1.3 & 0.096 & 0.000 & 0.059 & 0.104 & 0.058 & 0.072 & 0.023 & 0.012 & 0.030 & 0.011 \\
1.4 & 0.049 & 0.000 & 0.054 & 0.053 & 0.082 & 0.186 & 0.003 & 0.015 & 0.040 & 0.983 \\
1.5 & 0.114 & 0.000 & 0.065 & 0.100 & 0.063 & 0.102 & 0.018 & 0.000 & 0.000 & 0.037 \\
\bottomrule
\end{tabular}
}
\caption{Direct report generation performance comparison for IU X-RAY findings and impressions (Llama-3.2-90B-Vision-Instruct).}
\label{tab:llama-openi}
\end{table*}

\begin{table*}[htbp]
\centering
\resizebox{\textwidth}{!}{
\begin{tabular}{@{}l@{\hspace{2pt}}l@{\hspace{2pt}}l@{\hspace{2pt}}l@{\hspace{2pt}}l@{\hspace{2pt}}l@{\hspace{2pt}}l@{\hspace{2pt}}l@{\hspace{2pt}}l@{\hspace{2pt}}l@{\hspace{2pt}}l@{\hspace{2pt}}}
\toprule
\textbf{Experiment} & \multicolumn{4}{c}{\textbf{Lexical metrics}} & \multicolumn{6}{c}{\textbf{Clinical Efficacy Metrics}} \\
\cmidrule(lr){2-5} \cmidrule(l){6-11}
& \textbf{BLEU-1} & \textbf{BLEU-4} & \textbf{ROUGE-L} & \textbf{METEOR} & \textbf{Pos F1} & \textbf{Pos F1@5} & \textbf{RadGraph} & \textbf{Neg F1} & \textbf{Neg F1@5} & \textbf{Hall. Prop.} \\
\midrule
\multicolumn{11}{c}{\textsc{Chexpert Plus (Findings)}} \\
\midrule
1.1 & 0.166 & 0.006 & 0.175 & 0.147 & 0.214 & 0.290 & 0.092 & 0.096 & 0.153 & 0.419 \\
1.3 & 0.079 & 0.004 & 0.100 & 0.088 & 0.230 & 0.266 & 0.069 & 0.069 & 0.180 & 0.339 \\
1.4 & 0.022 & 0.003 & 0.062 & 0.029 & 0.247 & 0.355 & 0.006 & 0.042 & 0.059 & 0.097 \\
1.5 & 0.152 & 0.005 & 0.173 & 0.131 & 0.261 & 0.285 & 0.092 & 0.109 & 0.170 & 0.355 \\
\midrule
\multicolumn{11}{c}{\textsc{Chexpert Plus (Impression)}} \\
\midrule
1.1 & 0.073 & 0.002 & 0.106 & 0.072 & 0.152 & 0.221 & 0.013 & 0.003 & 0.007 & 0.173 \\
1.3 & 0.086 & 0.002 & 0.100 & 0.084 & 0.174 & 0.238 & 0.013 & 0.031 & 0.081 & 0.058 \\
1.4 & 0.091 & 0.000 & 0.087 & 0.071 & 0.190 & 0.281 & 0.005 & 0.000 & 0.000 & 0.985 \\
1.5 & 0.067 & 0.001 & 0.102 & 0.069 & 0.154 & 0.209 & 0.012 & 0.003 & 0.007 & 0.157 \\
\bottomrule
\end{tabular}
}
\caption{Performance comparison for Chexpert Plus findings and impressions (Llama-3.2-90B-Vision-Instruct).}
\label{tab:llama-chexpert}
\end{table*}

\begin{figure*}[t]
  \centering
\begin{subfigure}[b]{0.45\textwidth}
    \includegraphics[width=\textwidth]{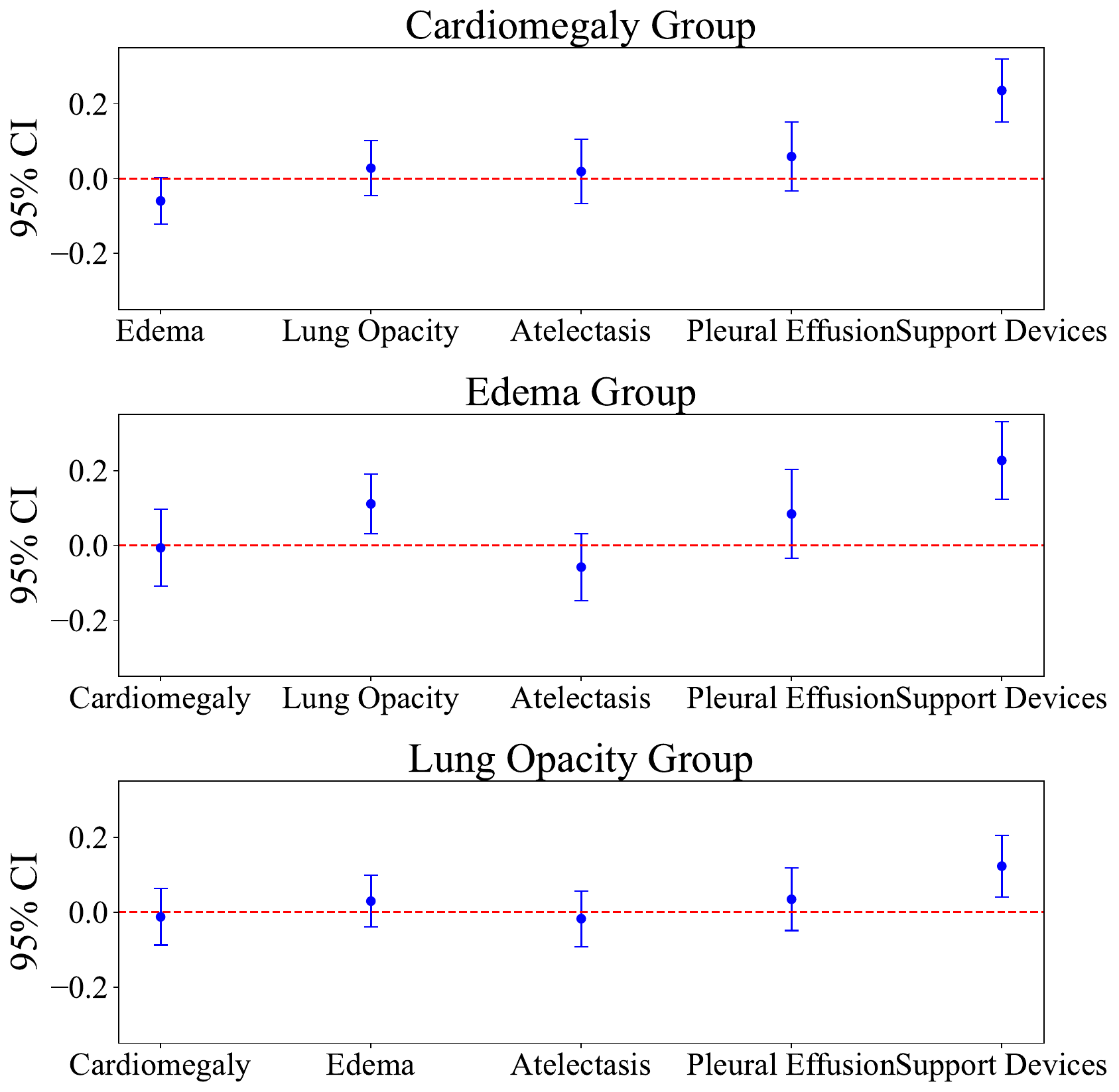}
    \label{fig:ci-first}
  \end{subfigure}
  \begin{subfigure}[b]{0.45\textwidth}
    \includegraphics[width=\textwidth]{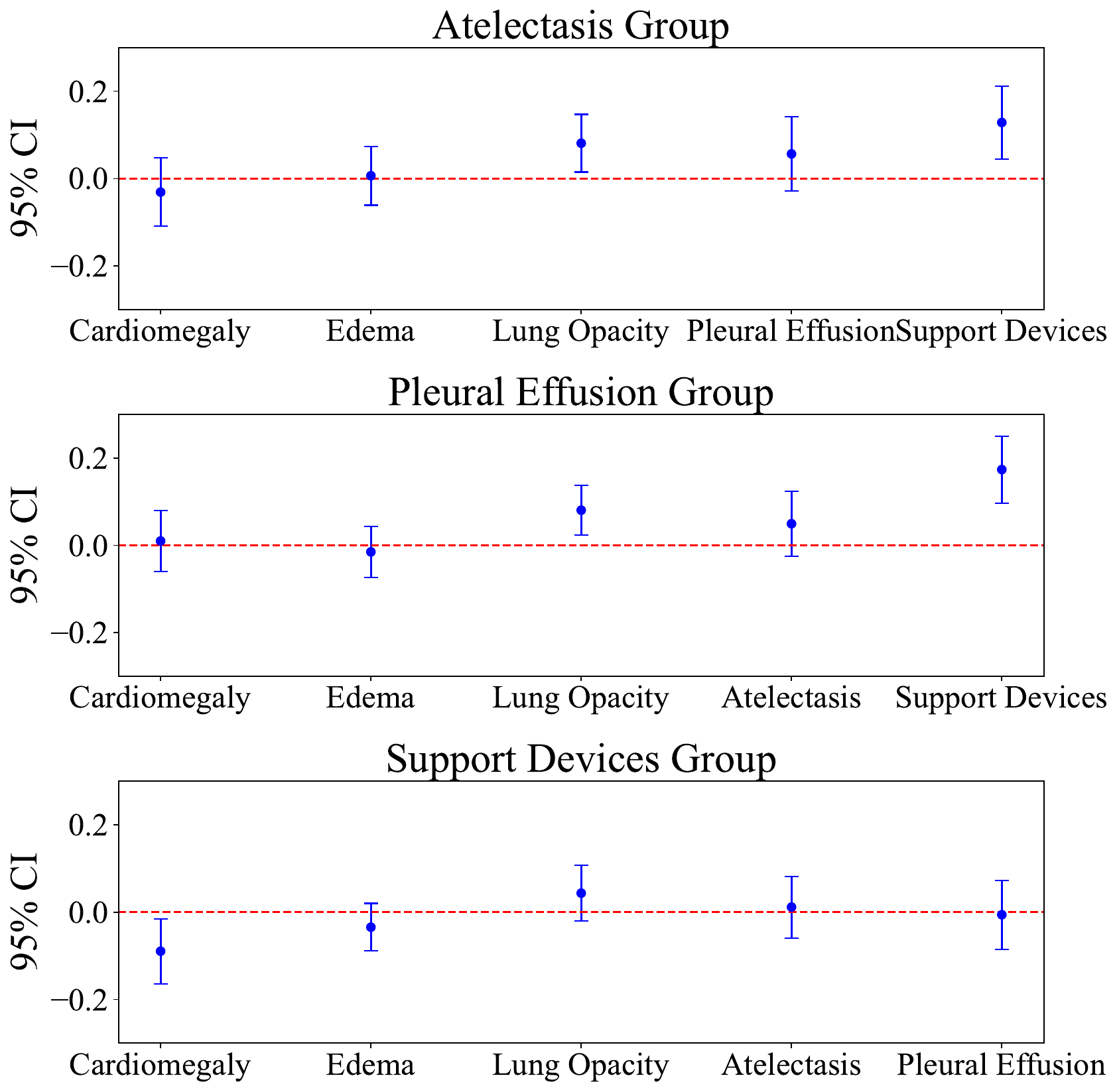}
    \label{fig:ci-4v}
  \end{subfigure}
  \captionsetup{skip=2pt}
  \caption{95\% Bootstrap confidence interval of top 6 conditions in our sample for GPT-4-vision-preview.}
  \label{fig:ci-4v}
\end{figure*}

\begin{table*}[h]
\centering
\footnotesize
\setlength{\tabcolsep}{1pt} 
\renewcommand{\arraystretch}{0.75} 
\resizebox{\linewidth}{!}{ 
\begin{tabular}{@{}>{\raggedright\arraybackslash}p{1.8cm}>{\raggedright\arraybackslash}p{7.5cm}>{\raggedright\arraybackslash}p{6cm}@{}} 
\toprule
\textbf{Experiment}  & \textbf{Findings}  & \textbf{Impression} \\ \midrule
GroundTruth & 
Single AP view of the chest provided. Interval increase in opacification of the right hemithorax with unchanged rightward shift of midline structures. The left lung is clear. Hilar and cardiomediastinal contours are obscured. &
Worsening collapse of the right lung from \_\_\_ with unchanged rightward shift of mediastinal structures. \\
\midrule
Basic &
The cardiomediastinal silhouette is severely enlarged. There is opacification of the right hemithorax with loss of visualization of the right heart border; this may be attributable to a large pleural effusion, consolidation, or a combination of both. The left lung field remains clear with normal aeration, without evidence of focal consolidation. The trachea is midline. There is no evidence of pneumothorax on the left. The osseous structures of the visualized portion of the thorax appear intact. Nasogastric tube is in place with its tip projecting over the gastric bubble. &
1. Severe cardiomegaly. 2. Opacification of the right hemithorax consistent with a large pleural effusion and/or consolidation. 3. No pneumothorax or acute bony abnormality identified. 4. The presence of a nasogastric tube with an appropriately situated tip. Further investigation with a CT scan may be considered to differentiate between effusion and consolidation and to assess for any additional intrathoracic pathology. Clinical correlation is recommended.\\
\midrule
+Indication &
The cardiac silhouette appears enlarged. There is an opacity in the right upper lobe consistent with lobar collapse and volume loss as evidenced by crowding of adjacent rib spaces. There are also patchy infiltrates seen in the right lower lobe. The remaining lung fields are clear without evidence of pleural effusion. The bony thorax and soft tissues appear unremarkable. &
Right upper lobe and right lower lobe collapse with an enlarged cardiac silhouette. These findings could be consistent with lung atelectasis secondary to obstruction, possibly due to an underlying mass or mucous plugging, and heart failure should be considered as a differential for cardiomegaly. Further diagnostic evaluation and correlation with clinical findings are recommended. \\
\midrule
+Instruction &
- Enlarged Cardiomediastinum is observed. - Cardiomegaly is present. - Evidence of Lung Opacity is seen. - Extensive consolidation is apparent. - There is Pleural Effusion. - Support devices in the form of a central venous catheter are noted. - No definite pneumothorax, pleural lesions, or fractures are identified.&
- Cardiomegaly with associated enlargement of the cardiomediastinum. - Significant lung opacity indicating consolidation, likely due to an infectious or inflammatory process, necessitating further clinical correlation. - Pleural effusion is present. - Central venous catheter in situ. - Absence of pneumothorax or significant pleural disease.\\
\midrule
CoT &
The cardiomediastinal silhouette is enlarged due to cardiomegaly. There are findings suggesting the presence of pulmonary edema, as evidenced by the perihilar alveolar opacities bilaterally, which may extend to the periphery. No evidence of atelectasis. There is a presence of pleural effusion and support devices, including endotracheal tube above the carina and a gastric tube in the stomach. No lung masses or nodules are identified. No definite pneumothorax or pneumoperitoneum.&
Cardiomegaly with signs of pulmonary edema and pleural effusion. Supportive devices in situ. No evidence of atelectasis, pneumothorax, or acute bony injury.\\
\midrule
Few-shot &
AP supine and lateral decubitus views of the chest demonstrate extensive consolidation involving the right hemithorax with associated volume loss. Evidence of air bronchograms within the consolidation suggests an alveolar filling process, such as pneumonia. There is near-complete opacification of the right lung, with the trachea deviated towards the right side, indicative of volume loss. No visible pneumothorax or significant pleural effusion is noted on the limited view available. The left lung is hyperinflated, likely compensatory. There is a loss of definition of the right heart border and diaphragm silhouette in keeping with the adjacent consolidation. The left hemidiaphragm and heart borders are normal. Osseous structures appear intact but are poorly visualized due to overlying opacity.&
Extensive right lung consolidation with volume loss, likely representing a severe pneumonia process. No clear evidence of pleural effusion or pneumothorax on the available projections. Compensatory hyperinflation of the left lung. Clinical correlation and possibly further imaging, such as a CT scan, are recommended for a comprehensive assessment.\\
\bottomrule
\end{tabular}
}
\caption{Comparison of generated reports across different prompting strategies for one study.}
\label{tab:rad}
\end{table*}

%---------------------------------------------end-------------------

\end{document}